\documentclass[journal, 10pt]{IEEEtran}
\usepackage[utf8]{inputenc}

\usepackage{amsmath,amssymb,amsfonts,amsthm}
\usepackage{algorithmic}
\usepackage{algorithm}
\usepackage{graphicx}
\usepackage{xcolor}
\usepackage{mathtools}
\usepackage{multirow}
\usepackage{dblfloatfix}
\usepackage{tikzpagenodes}

\DeclareMathOperator*{\In}{In}
\DeclareMathOperator*{\inam}{in}
\DeclareMathOperator*{\cont}{cont}
\DeclareMathOperator*{\Cont}{Cont}
\DeclareMathOperator*{\Tr}{Tr}

\DeclareMathOperator*{\val}{val}
\DeclareMathOperator{\Distance}{Distance}
\DeclareMathOperator{\semsim}{sim}
\DeclareMathOperator{\IC}{IC}
\DeclareMathOperator{\LS}{LS}
\DeclareMathOperator{\wt}{wt}
\DeclareMathOperator{\Dist}{Dist}
\newcommand{\defeq}{\vcentcolon=}

\usepackage{booktabs}
\newcommand{\tabitem}{~~\llap{\textbullet}~~}

\newcommand{\etal}{\textit{et al. }}

\theoremstyle{remark}
\newtheorem{thm}{Theorem}

\usepackage[style=ieee, backend=biber]{biblatex}
\title{Engineering Semantic Communication: A Survey}

\author{
    Dylan Wheeler, \IEEEmembership{Graduate Student Member, IEEE} and Balasubramaniam Natarajan, \IEEEmembership{Senior Member, IEEE}

    \thanks{D. Wheeler and B. Natarajan are with the Mike Wiegers Department of Electrical and Computer Engineering at Kansas State University (email: dylan84@ksu.edu)}
}

\begin{document}

\maketitle

\begin{tikzpicture}[remember picture,overlay]
    \node[align=center] at ([yshift=1em]current page text area.north) {This work has been submitted to the IEEE for possible publication. Copyright may be\\ transferred without notice, after which this version may no longer be accessible.};
\end{tikzpicture}%

\begin{abstract}
    As the global demand for data has continued to rise exponentially, some have begun turning to the idea of semantic communication as a means of efficiently meeting this demand. Pushing beyond the boundaries of conventional communication systems, semantic communication focuses on the accurate recovery of the \textit{meaning} conveyed from source to receiver, as opposed to the accurate recovery of transmitted symbols. In this survey, we aim to provide a comprehensive view of the history and current state of semantic communication and the techniques for engineering this higher level of communication. A survey of the current literature reveals four broad approaches to engineering semantic communication. We term the earliest of these approaches classical semantic information, which seeks to extend information-theoretic results to include semantic information. A second approach makes use of knowledge graphs to achieve semantic communication, and a third utilizes the power of modern deep learning techniques to facilitate this communication. The fourth approach focuses on the significance of information, rather than its meaning, to achieve efficient, goal-oriented communication. We discuss each of these four approaches and their corresponding studies in detail, and provide some challenges and opportunities that pertain to each approach. Finally, we introduce a novel approach to semantic communication, which we term context-based semantic communication. Inspired by the way in which humans naturally communicate with one another, this context-based approach provides a general, optimization-based design framework for semantic communication systems. Together, this survey provides a useful guide for the design and implementation of semantic communication systems.
\end{abstract}


\vspace{-0.4cm}

\section{Introduction}
\label{sec_intro}

\begin{figure}[htbp]
    \centering
    \includegraphics[scale = 0.49]{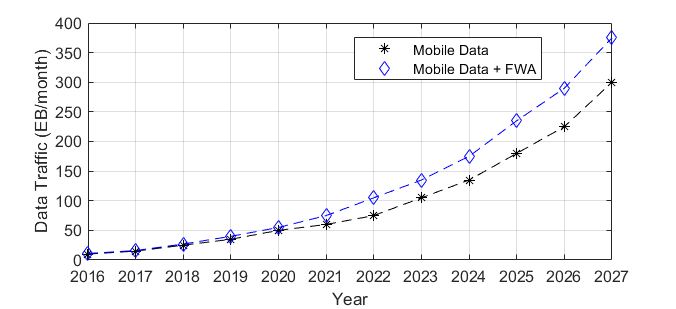}
    \caption{Global monthly traffic predictions from mobile data and fixed wireless access (FWA) \cite{ericsson_report}}
    \label{fig_traffic}
\end{figure}

As 5G continues to roll out across the globe, the world of wireless communications continues to expand and grow. According to a report published by Ericsson in November of 2021, the monthly global data traffic is predicted to grow exponentially over the next five years \cite{ericsson_report} (see Figure \ref{fig_traffic}). Recently, the circumstances imposed by the ongoing COVID-19 pandemic have sparked a movement of an increasing number of people choosing to telecommute for work \cite{telework_report}. This will no doubt accelerate global traffic growth even further. 
This unprecedented growth is accompanied by an array of new use cases for wireless networks. As defined by the 3rd Generation Partnership Project (3GPP), the 5G network is focused on supporting three main use cases, namely (1) enhanced mobile broadband (eMBB),  (2) massive machine-type communication (mMTC), (3) and ultra-reliable low-latency communication (URLLC) \cite{jiang_6G}. eMBB is aimed at providing enhanced services to traditional users of the mobile network, focusing on increased throughput and connection density. mMTC is a paradigm that is designed for a large network of devices each transmitting a relatively small amount of data, e.g. a large wireless sensor network. URLLC is needed when communications are critical and extremely time-sensitive, e.g. when performing remote surgery, or during complex manufacturing processes.

From these observations, we anticipate two trends. First, global data traffic will continue to increase at an exponential rate. As all data communication requires some amount of energy for transmission, this will translate to an exponential growth in overall power consumption of wireless networks. In a world where we all are under increasing pressure to reduce consumption and create more sustainable infrastructure, this trend presents a grand challenge indeed. Second, the decision by the 3GPP to define three specific use cases of the 5G network points to another trend, which is the growing heterogeneity of the wireless network. Communication across the network is increasingly diverse, with a wide variety of use cases and application-specific requirements and constraints. To be able to meet the growing demand for data in a sustainable way, the crucial question is this: \textit{how can we communicate more efficiently over an increasingly heterogeneous network?}

Considering the recent, widespread success demonstrated by artificial intelligence (AI) and machine learning (ML), we believe that a promising approach to address this challenging question is to develop more intelligent communication systems, and specifically, \textit{semantic communication} systems. In their pioneering work, Shannon and Weaver defined three fundamental communication problems \cite{shannon_weaver}:
\begin{enumerate}
    \item[A.] How accurately can the symbols of communication be transmitted? (The technical problem.)
    \item[B.] How precisely do the desired symbols convey the desired meaning? (The semantic problem.)
    \item[C.] How effectively does the received meaning affect conduct in the desired way? (The effectiveness problem.)
\end{enumerate}

\begin{figure*}[htbp]
    \centering
    \includegraphics[scale = 0.57]{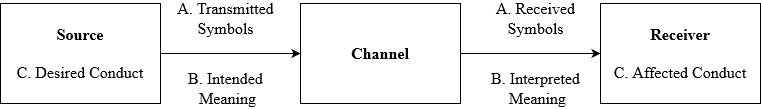}
    \caption{General model of communication with components of the three fundamental communication problems}
    \label{fig_3probs}
\end{figure*}

The general model of a communication system is given in Figure \ref{fig_3probs}, with these three problems superimposed. 
All communication systems today operate at the technical level, i.e., trying to recover transmitted symbols at the receiver as accurately as possible, with no regard for what the symbols mean. Shannon himself stated in his seminal 1948 paper that ``the fundamental problem of communication is that of reproducing at one point either exactly or approximately a message selected at another point. Frequently the messages have meaning... These semantic aspects of communication are irrelevant to the engineering problem" \cite{shannon}. 

Clearly, Shannon's view on communication was visionary, and it enabled the extraordinary growth that we have seen in communication systems to this day. However, when subscribing to this view today, we are limited in our options of how to address the previously discussed challenge. If we operate solely at the technical level, an increased demand translates directly to an increased consumption of resources, in the form of either power and/or bandwidth. Increased power consumption is exactly what we are attempting to avoid, and bandwidth is increasingly scarce. Existing usable bandwidths are extremely crowded \cite{fcc_spec_alloc}, and there are known issues when operating at higher frequencies (high attenuation, variability, etc.). Recent advances in powerful technologies such as beam forming and massive MIMO \cite{mimo_beamform} can serve as a temporary solution to this problem, but in the face of exponential demand growth, even these will eventually fall short. 

Instead of engaging in the unsustainable pursuit of increasing resource consumption to meet demands, some have suggested a turn toward operating at the second level of communication instead, namely the semantic level \cite{strinati_toward6G}. If such a communication system is achievable, could it enable more efficient communication? Intuitively, the rationale is sound. To illustrate, let us consider two scenarios of human-to-human communication. In both scenarios, the speaker is trying to teach the listener how compute the area of a circle. In scenario 1, let's say that the listener is someone at least vaguely familiar with geometric concepts, while in scenario 2, the listener is a young child. In both scenarios, the semantic problem is exactly the same; namely, convey what it means to compute the area of a circle. However, the technical problems will likely be very different. In scenario 1, communication may well be very efficient, and perhaps all that is needed is to provide the formula $A = \pi r^2$. In scenario 2, much more information will be required from a technical perspective. To enable understanding of the formula, the speaker would first need to clarify what each piece means, e.g., $r$ represents the radius of the circle, which is the distance from the center of the circle to the edge of the circle, and so on. The speaker would likely need to speak slower and repeat some key points to fully convey the meaning.

What influences the speaker's approach in either scenario is the difference in prior knowledge bases of the listeners, and the \textit{speaker's knowledge} of this difference. Note that, as a trivial solution, the speaker can use the approach in scenario 2 in both cases, i.e., the speaker can always assume the worst case scenario (no prior knowledge base) and thoroughly explain every aspect of the problem. Let's call this the semantics-agnostic approach. While this will ensure complete conveyance of meaning in both scenarios, it is clear that the speaker is wasting resources (time, energy) by speaking to the listener in the first scenario as if they were a child. This key intuition is the driving force behind the push toward semantically-aware communication systems.

We envision semantic communication as a paradigm shift for future communication systems that embodies a natural progression based on the three communication problems outlined by Shannon and Weaver. By using semantic systems to communicate intelligently, we believe that this technology can meet the challenge posed by rising data demands. Despite its early beginnings in 1952 with Carnap and Bar-Hillel's work on a theory of semantic information \cite{carnap}, the body of literature regarding semantic communications is quite small. At the time of writing, a search of the phrase ``semantic communication'' returns only 42 results in IEEEXplore. However, 16 of these results were published in the past year alone, indicating a surge of interest in the field. Therefore, we feel that a paper surveying these works will be invaluable as the field develops out of infancy. In this paper, we attempt to provide as clear a picture as possible of the current state of semantic communications. Next, we discuss some of the recent works providing their own summaries and visions of semantic communications. We then offer a brief discussion of what it means to define semantics, followed by a presentation of different approaches found in the literature, and finish with our own take on a natural approach to the problem of engineering\footnote{To avoid confusion, we note that by ``engineering,'' we refer to ``the action of working artfully to bring something about.'' \cite{oed_engineering_def} Most of the works described in this survey entail theoretical developments, rather than physical systems.} semantic communication. 

\subsection{Related Work}
\label{subsec_rel_work}

There have been several published works that attempt to provide a vision of what semantic communication might look like, and even fewer that attempt to survey the young field. While most of these have come about in recent years, one of the earliest examples was published in 1992 by Ouksel and Naiman \cite{ouksel_92}, in which they discuss a semantic communication protocol in heterogeneous database systems. They argue that a semantic communication protocol provides a more flexible vehicle of communication and can support effective conflict resolution. Although they refer to this protocol as semantic communication, their ideas align much more with the concept of the \textit{semantic web}, which was first introduced by Berners-Lee \textit{et al.} in 2001 \cite{berners_semantic_web}. While related, the idea of the semantic web is different from what we refer to as semantic communication. The primary goal of the semantic web is to make the information contained in pages on the internet machine-interpretable. This field has been well-studied over the years since its inception. The semantic web can be thought of as an attempt to achieve the second (semantic) level of communication between the web (source) and a machine (receiver), and thus represents a particular case of general semantic communication. In this paper, we aim to keep our discussion centered on this more general problem.

One of the earliest works providing a vision for addressing the semantic communication problem was given by Rodoplu \textit{et al.}  in 2007 \cite{rodoplu_07}. They introduce their idea of a semantic domain, which includes semantic ``atoms'' and corresponding atomic operations that act on the atoms to generate objects. They then provide their vision of how one might characterize semantic information, making the important observation that the same object might have different semantic information measures in different domains. The vision provided in \cite{rodoplu_07} leans on the idea of using knowledge graphs to represent prior knowledge bases at both the transmitter and receiver, falling within the realm of what we refer to as knowledge graph-based semantic communication, which is the focus of section \ref{subsec_ontol}.

Another vision article from 2013 \cite{juba_2013} focuses specifically on the problem of semantic misunderstanding. An illustrative example given is the failure of the Mars Climate Orbiter, in which two collaborating teams of engineers were working in different unit systems (imperial and metric), leading to misunderstandings and failure of the mission. The article focuses more on the effectiveness problem rather than the semantic problem, and argues that the key theoretical notion for successful communication is the presence of \textit{sensing}; which is described as feedback to the source indicating successful communication. The vision presented in this article describes agents in a communication system that can learn to achieve successful communication (in the effectiveness sense) despite some initial knowledge mismatch between source and receiver.

In \cite{juang_2011}, a brief history is provided on the quantification and transmission of information and intelligence. In addition to providing an informative summary on the history of communication techniques and theory, the article raises some important challenges for the design of future intelligent communication, namely:
\begin{itemize}
    \item Can the formulation of channel capacity include a function of significance?
    \item How do we define error in the transmission of intelligence?
    \item How can the code set and signal shaping be defined to support optimal transmission of intelligence?
    \item How can the receiver be design to optimally accomplish reception of intelligence?
\end{itemize}
By interpreting the phrase ``transmission of intelligence'' as the nearly synonymous term of semantic communication, these represent some of the grand challenges of practically achieving this higher level of communication. One possible avenue is proposed to address these challenges, which suggests the use of Bayes' decision theory to quantify the ``significance'' of information.

Building on the idea of information significance, in \cite{uysal_semcomm_2021} Uysal \textit{et al.} envision semantics to mean just that: the significance of information as opposed to its meaning. They argue for an extensive cross-layer optimization of the end-to-end communication system, in a self-described \textit{radical departure} from the well-established way of assessing communication networks. The key idea is that this optimization will yield semantic communication by ``the provisioning of the right piece of information to the right point of computation (or actuation) at the right point in time." The development of semantic measures is called for to quantify what information is ``right'' such that the system can be properly optimized. The significance-focused view on semantics is also advocated for in \cite{kountouris_2021}, which describes a vision of goal-oriented semantic communication, essentially combining the semantic and effectiveness problems. In addition to outlining this vision, utilizing these ideas is shown to greatly reduce robotic actuation error in a provided example. This view on semantics has much in common with our proposed vision in Section \ref{sec_context}, and is discussed in further detail in later sections.

A recent article outlining nine challenges in AI for 6G communications \cite{tong_21} points out the capabilities of recent learning techniques as a potential enabler of semantic communication. The article outlines two challenges which must be met to develop semantic communications, the first of which is the mathematical foundation of semantic communications; while some attempts have been made at defining a mathematical framework on which to build semantic communication, we are still lacking a comprehensive theory. The second challenge is the structure of semantic communication systems, which is posed as a problem of choosing between a general deep neural network (DNN) or further exploring other structural levels of communication to facilitate semantic communication. Another vision article promoting the use of ML in semantic communication for 6G networks is \cite{strinati_21}. This extensive article provides a complete view on the current vision for 6G networks, while also discussing details of both semantic and goal-oriented (effective) communications, with various examples and applications given for each. They then provide their vision of the 6G network as incorporating online learning-based communication and control. In contrast to our goal of providing a clear and comprehensive view of the field of semantic communications, \cite{strinati_21} focuses on ML and applications of semantic communication. Similar to \cite{tong_21} and \cite{strinati_21}, \cite{luo_2022} provides an overview of end-to-end semantic communications based on DL. The discussion is broken into semantic communications for different modalities, such as text, image and speech. Use cases, including internet of things (IoT) networks and smart factories are discussed, and open issues are presented.

\begin{figure*}
    \centering
    \includegraphics[scale = 0.70]{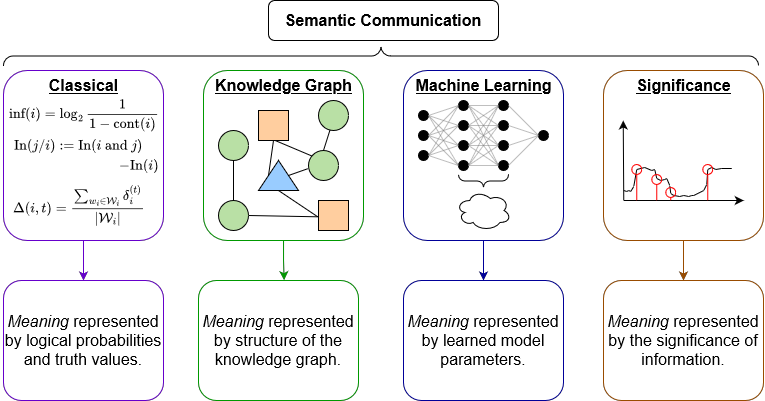}
    \caption{The four existing approaches to semantic communication}
    \label{fig_overview_semcomm}
\end{figure*}

The work that is perhaps the most similar to that presented throughout this survey is the recent review published by Lan \textit{et al.} \cite{lan_21}, in which the authors review principles of semantic communication, discuss existing system architecture designs, and provide an overview of designing semantic communication systems based on knowledge graphs. Their discussion is divided into the categories of human-human (H2H), human-machine (H2M), and machine-machine (M2M) communication, with example applications provided for each. While sharing many similarities, \cite{lan_21} differs from the work here in a few ways. First, the discussion contained in \cite{lan_21} is application-centric, while we aim to focus more on the general techniques and foundations of semantic communication. This leads to a natural contrast in presentation; \cite{lan_21} breaks the discussion into H2H, H2M, and M2M based on the \textit{application} of semantic communication, while we partition topics in our discussion based on the \textit{definition} of semantics in communication. We believe this provides a clearer view of the way in which semantic communications are thought about today. Furthermore, unlike \cite{lan_21}, we discuss the classical approaches to quantify semantic information, in order to provide perspective on how the field has evolved since its inception.

As mentioned above, the purpose of this survey is to provide a clear picture of the history and current state of semantic communications, with the hope that it will be a useful guide to those wishing to pursue research within this exciting field. To that end, the rest of this paper will be organized as follows. In Section \ref{sec_def_sem}, we offer some perspective on the difficulties of defining semantic communication. There have been many different ideas on just how to do this, and we attempt to group these in a natural way. Based on this grouping, Sections \ref{subsec_classic}-\ref{subsec_right} will review works that fall into each category. In Section \ref{subsec_classic}, an overview of classical semantic information theory is provided. Section \ref{subsec_ontol} reviews works falling under the category of knowledge graph-based semantic communication, which has traditionally been the most common approach. ML-based semantic communication is considered in Section \ref{subsec_ml}. This approach has seen a surge of activity in recent years, and is a promising approach moving forward. Section \ref{subsec_right} goes into more depth on the recently proposed approach of treating semantics as the significance of information. Building off of this idea, in section \ref{sec_context} we present our vision on an alternate approach to semantic communication, which emphasizes context as the core component. Section \ref{sec_conclude} concludes the paper and offers some future research directions.

\section{Overview of Semantics in Communication}
\label{sec_def_sem}

The ambiguity of the word ``semantics'' brings with it an inherent difficulty when attempting to provide a definition. Indeed, this is an issue that has drawn the attention of engineers and philosophers alike. In \cite{juba_book}, a brief discussion regarding the philosophical context of semantic communication is provided. There, it is noted that the idea that ``communication must be considered as a means to an end'' was first brought about by Dewey in 1925 \cite{dewey_1929}, and later ``brought to the forefront of philosophy'' by Wittgenstein in 1953 \cite{wittgenstein_1953}. In this work, Wittgenstein fills a short book, organized into a continuous flow of philosophical remarks, with his thoughts and reflections on the fundamental aspects of language. Clearly then, having been at the center of a great amount of philosophical thought, the definition of semantics is a complex one.

However, as we are interested in the engineering of semantic communication, this definition is vital. Without it, we are left blind when attempting to create systems which may achieve this higher level of communication. The solution to any engineering problem requires full understanding of the problem itself, including any simplifications, assumptions, and constraints posed by the problem. Therefore, we dedicate this section to a brief introduction to the different engineering problems that have been posed when attempting to create semantic communication. Inherent in each is the way in which this abstract idea of semantics is defined, which leads to different implications and solutions. Figure \ref{fig_overview_semcomm} provides a high-level view of these approaches and the ways that meaning is represented by each.

As previously mentioned, in their 1952 paper Carnap and Bar-Hillel attempt to outline a Theory of Semantic Communication \cite{carnap} as a direct response to Shannon's then-recently published ``A Mathematical Theory of Communication." The core of their work is to base information measures around \textit{logical} probabilities rather than the \textit{statistical} probabilities which underlie what we now call Information Theory. This definition of semantics is concerned with the so-called logical truth of a statement, from which information measures can be derived. We classify this definition and its derivatives \textit{classical semantic information}. However, it is noted in \cite{carnap} that this definition of ``semantic information is a concept more readily applicable to psychological and other investigations than its communicational counterpart." Regardless, there are still some important works that embrace this idea of semantic information for the engineering problem.

Perhaps the most pervasive method of defining semantics throughout the literature is to do so by using some sort of structured knowledge base. This structured knowledge base can take on many names, such as ``semantic network'' \cite{delgadofrias}, ``taxonomy'' \cite{rada} , ``ontology'' \cite{mertoguno}, and others. All of these essentially refer to the same idea, which is to use a graph structure, or \textit{knowledge graph} to represent knowledge in the system  \cite{ji_kg_survey}. Hence, we refer to these techniques as \textit{knowledge graph-based semantic communication}. Having close ties to the semantic web, it is clear why this approach is popular. By defining knowledge over a graph, it is relatively straightforward to define metrics of ``semantic similarity,'' which can then be analyzed using well-developed graph theory techniques. Furthermore, recent work on graph neural networks brings about the opportunity to incorporate modern learning techniques over such graphs  \cite{graph_neuralnets_survey}.

Another prevalent approach that is seeing a surge of interest is the idea of using ML techniques to ``learn'' the semantics of a problem. Akin to model-based vs. data-driven approaches to general inference problems (see \cite{shlezinger_modelbaseddl} for more on this), predefined knowledge graphs impose model-based semantics on the problem at hand while ML methods use data to determine these semantics. Borrowing techniques from natural language processing (NLP) and computer vision, deep networks can be taught to communicate in the most efficient manner while preserving semantic content \cite{xie_deepsc}. Similarly, by implementing reinforcement learning (RL) methods, these networks can be refined over time and even adapt to dynamic changes in the communication problem. We refer to this approach as \textit{machine learning-based semantic communication}. Initial studies examining this approach to semantic communication over several modalities have emerged in recent years \cite{xie_deepsc,xie_lite_deepsc,xie_deepsc_speech,xie_deepsc_vqa}.

Finally, a fourth definition which differs significantly from those previously described is the one first mentioned in Section \ref{subsec_rel_work}: the idea of semantics as the ``significance'' of information; we call this \textit{significance-based semantic communication}. While classified as an approach toward semantic communication, this approach essentially addresses the third level of communication problem: the \textit{effectiveness} problem. Rather than concern ourselves with the meaning of a message, advocates of this approach call for communication of the \textit{right} information. Of course, what information is ``right'' will depend on the application and the desired outcome, therefore leading to the idea of \textit{effective} or \textit{goal-oriented communication}. This approach lends itself well to machine-machine communication, in which we are less concerned with the conveyance of ``meaning'' and more concerned with what the system accomplishes. One well-studied way of quantifying what information is ``right'' is the popular age of information metric  \cite{age_of_info_survey}, which relates to generating and delivering information at the right time. By defining other metrics to quantify what is ``right'' for the problem at hand, a joint optimization can be carried out to achieve optimal communication.

These four approaches to defining semantics roughly partition the previous research regarding semantic communication. Based on this analysis of the current literature, we broadly define semantics as \textit{any definition of information or the transfer thereof that considers something beyond the statistical nature of the symbols used to represent that information.} This view unites the definitions discussed above, despite their differences in the ``something beyond'' that is considered by each. Of course, these definitions are not mutually exclusive. ML can be used to determine which information is ``right,'' just as classical semantic information metrics can be used to design and tune neural networks. Just as model-based deep learning (DL) incorporates both prior knowledge and data-driven techniques for inference, graph neural networks can be used to learn knowledge graphs for semantic communication. However, by treating each of these definitions individually, a complete picture is given of the current state of semantic communication. In the next sections, we dive into each definition, the engineering approaches that come as a result, and some of their potentials and challenges.

\section{Classical Semantic Information}
\label{subsec_classic}

As mentioned earlier, Carnap and Bar-Hillel's paper \textit{An Outline of a Theory of Semantic Information} attempts to tackle the problem of engineering semantic communication through the quantification of semantic information \cite{carnap}. This quantification is the first step towards efficient semantic communication, as it allows us to consider ideas such as semantic compression and semantic error. Therefore, we begin our discussion with the ideas and methods related to classical semantic information. The works discussed in this section are summarized in Table \ref{tab_classic_summary}.

{\renewcommand{\arraystretch}{1.5}
\begin{table*}[htbp]
    \centering
    \caption{Summary of Works in Classical Semantic Information}
    \begin{tabular}{|c|c|c|}\hline
    \multirow{3}{*}{Theory of Weakly Semantic Information}     & Carnap \& Bar-Hillel {\csname @fileswfalse\endcsname\cite{carnap}} & Semantic information using logical probabilities\\\cline{2-3} 
                                                               & Bao \textit{et al.} {\csname @fileswfalse\endcsname\cite{bao}} & Definition of semantic entropy, corresponding theorems\\\cline{2-3}
                                                               & Basu \etal {\csname @fileswfalse\endcsname\cite{basu}} & Semantic compression\\\hline 
   Theory of Strongly Semantic Information  & Floridi {\csname @fileswfalse\endcsname\cite{floridi}} & Extension of TWSI using truth values    \\\hline 
    Truthlikeness   &  D'Alfonso {\csname @fileswfalse\endcsname\cite{dalfonso}}   & Semantic information based on truthlikeness                        \\\hline 
    \end{tabular}
    \label{tab_classic_summary}
\end{table*}
}

\subsection{Theory of Weakly Semantic Information}

Following the convention of \cite{floridi}, we will refer to the theory laid out by Carnap and Bar-Hillel as the Theory of Weakly Semantic Information (TWSI), for reasons that will be discussed later. TWSI is defined over a language system $\mathcal{L}_n^\pi$ which contains $n$ ``things'' (or individuals) and $\pi$ primitive one-place predicates (descriptors). An \textit{atomic} sentence is said to consist of a single predicate describing a single thing, while a \textit{molecular} sentence is formed from two or more atomic sentences joined with some logical connective, including: or, and, if...then, if and only if. Any sentence can either be logically true, logically false, or logically indeterminate. Furthermore, \textit{logical relations} are defined. For example, for sentences $i$ and $j$, we have $i$ \textit{logically implies} $j$ defined to mean that ``if $i$ then $j$'' is logically true. A \textit{state description} is a sentence in which each of the $\pi$ predicates is specified for each of the $n $ individuals; thus completely specifying all aspects of the universe. Common set notation can be used to talk about ``classes'' of entities within the universe. For example, \cite{carnap} describes a system consisting of three individuals, $\{a,b,c\}$, and two binary predicates, young or old ($Y$ or $O$) and male or female ($M$ or $F$). Then an example of a state description could be given as ``($a$ is $F$ and $Y$) and ($b$ is $M$ and $Y$) and ($c$ is $M$ and $O$)." Some other possible states are given below in Table \ref{tab_twsi_example}.

{\renewcommand{\arraystretch}{1.2}
\begin{table}[htbp]
    \centering
    \caption{Some states of an example universe}
    \begin{tabular}{|c|c|c|c|c|c|}
    \hline
    \multicolumn{2}{|c}{$a$} & \multicolumn{2}{|c}{$b$} & \multicolumn{2}{|c|}{$c$}\\\hline
        $M/F$ & $Y/O$ & $M/F$ & $Y/O$ & $M/F$ & $Y/O$\\\hline
        $F$&$Y$&$M$&$Y$&$M$&$O$\\\hline
        $F$&$O$&$M$&$O$&$M$&$O$\\\hline
        $M$&$Y$&$M$&$Y$&$M$&$Y$\\\hline
        $F$&$O$&$M$&$Y$&$F$&$Y$\\\hline
    \end{tabular}
    \label{tab_twsi_example}
\end{table}
}

Note that, for binary predicates, a universe will consist of $2^{n\pi}$ possible state descriptions. Similar to the process in which Shannon developed entropy as a measure of information \cite{shannon}, Carnap and Bar-Hillel begin with requirements/axioms that semantic information (and its corresponding measures) must satisfy. Denoting the semantic information of a sentence as $\In(\cdot)$, the first axiom is given as
\begin{equation}
    \In(i) \textrm{ includes } \In(j) \iff i \textrm{ logically-implies } j,
    \label{eq_carnap_ax1}
\end{equation}
meaning that $i$ says everything that is said by $j$ (and possibly more) if and only if $j$ is implied by $i$ within the logical framework. For example, take $i = $ ($a$ is $F$ and $Y$) and $j = $ ($a$ is $F$). Clearly, $j$ is implied by $i$, and thus the semantic information of $i$ includes that of $j$ (though the converse is not true). This axiom requires us to treat information as a \textit{set} or \textit{class} of something; it is important to note that the \textit{amount} of information is arbitrary at the moment, and must be defined on this set. Some theorems are derived from this axiom, and the concept of \textit{relative information} is defined as
\begin{equation}
    \In(j/i) \defeq \In( i \textrm{ and } j) - \In(i)
    \label{eq_carnap_rel}
\end{equation}
where $\In(j/i)$ is again some set or class. Continuing the example above, we can see that $\In(j/i) = \emptyset$, since $\In(i \text{ and } j) = \In(i)$. However, $\In(i/j) \neq \emptyset$.

Based on (\ref{eq_carnap_ax1}) and (\ref{eq_carnap_rel}), a concept of the information of a sentence is defined, and is termed the \textit{content} of a sentence. Content is derived from what is called a \textit{content-element}, which is simply the negation of a state description. The content of a sentence $i$, denoted $\Cont(i)$, is then defined as the set of content-elements logically implied by $i$. Intuitively, $\Cont(i)$ can be thought of as the set of state descriptions in the overall state-space (i.e. universe described by our language system) that are \textit{eliminated} with knowledge of the sentence $i$. Take $i = $ ($a$ is $F$ and $Y$) as before. Out of the $2^6 = 64$ possible state descriptions in the universe, this sentence eliminates all 48 in which $a$ is not $F$ or $a$ is not $Y$, and $\Cont(i)$ is composed of the negations of all such state descriptions. Clearly then, a self-contradiction will ``say the most'' (by eliminating all state-descriptions) and a tautology will ``say the least'' (by eliminating no state-descriptions). Similarly, a complete state description can be thought of as carrying much information, since it eliminates all other state descriptions. Note that this notion of information has nothing to do with the truth of a sentence, which is a point we will revisit. With this idea of information in place, the primary question is addressed: how shall the \textit{amount} of information be defined? 

The amount of semantic information carried by sentence $i$ is denoted as $\inam(i)$, and the following requirements are stated:
\begin{align}
    \label{eq_inam_r1}
    \Cont(j) \subseteq \Cont(i) &\Rightarrow \inam(i) \geq \inam(j)\\
    \label{eq_inam_r2}
    \Cont(j) \subset \Cont(i) &\Rightarrow \inam(i) > \inam(j)\\
    \label{eq_inam_r3}
    \Cont(i) = \emptyset &\Rightarrow \inam(i) = 0
\end{align}
Note the subtle difference between $\In(\cdot)$ and $\inam(\cdot)$; $\In(\cdot)$ represents the information itself, while $\inam(\cdot)$ is used to quantify \textit{how much} information is given by $\In(\cdot)$ Using $\Cont(\cdot)$ as the information content of a sentence, these requirements are straightforward and make intuitive sense. By (\ref{eq_inam_r1}) and (\ref{eq_inam_r2}), a sentence containing all the information of another should have a greater than or equal amount of information, with equality only if the information carried is the same. By (\ref{eq_inam_r3}), the amount of information is zero if the information of a sentence is the empty set.

Based on these requirements, two measures are offered as the main contribution of \cite{carnap}. The first is termed the \textit{content-measure} of a sentence, denoted $\cont(\cdot)$ (different from $\Cont(\cdot)$), and is defined as any proper $m$-function of the negation of a sentence. We will not go into the details of what constitutes an $m$-function here (see \cite[Section 6]{carnap} for further reading), but suffice it to say that it satisfies (\ref{eq_inam_r1})-(\ref{eq_inam_r3}), and defines a measure taking values between 0 and 1, thus representing a logical probability measure. However, a problem arises with this measure regarding another intuitive requirement not yet stated, namely additivity. Just as in classic Information Theory, we would like the information measure of two \textit{independent} sentences to follow additivity, i.e. $\inam(i\textrm{ and }j) = \inam(i) + \inam(j)$ for $i,j$ independent; it is shown that $\cont(\cdot)$ does not satisfy this intuition \cite[Thm. 6-15]{carnap}. Thus, a second measure is proposed and is termed \textit{measure of information}, denoted by $\inf(\cdot)$, not to be confused with the infimum operator. This second measure is defined as
\begin{equation}
    \label{eq_carnap_inf}
    \inf(i) = \log_2 \frac{1}{1 - \cont(i)}.
\end{equation}
Observe that this measure is analogous to the classical information-theoretic definition of entropy, making use of the logical probability $\cont(i)$ instead of the statistical probability $p(i)$.

A comparison of these two measures is given in \cite{carnap}, and it is shown that both exhibit intuitively desirable properties, and likewise they both exhibit intuitively undesirable properties. The lack of additivity of $\cont(\cdot)$ is one example. Another is that $\inf(\cdot)$ lacks a counterpart to the property of $\cont(\cdot)$ stating that $\cont(i\textrm{ and }j) \leq \cont(i) + \cont(j)$. Thus, it is concluded that neither represent an ideal measure of semantic information, but rather that they both have specific strengths and weaknesses. 

\subsection{Theory of Strongly Semantic Information}

A problem with TWSI occurs when presented with a sentence that constitutes a contradiction, e.g., ``$i$ and not $i$.'' As mentioned above, under the definition of $\Cont(\cdot)$, this sentence would carry with it maximum semantic information. Intuitively, however, we know that a contradiction \textit{should} carry no information; it is obviously untrue and leaves the receiver no less informed than before the reception of the message. This ambiguity manifests itself in the mathematics of TWSI as well, and is known as the Bar-Hillel-Carnap Paradox (BCP). Therefore, in \cite{floridi} Luciano Floridi proposes that \textit{truth} lies at the root of this paradox, which can be solved by the incorporation of truthfulness considerations into TWSI.

The theory which follows is outlined in  \cite{floridi} and is deemed a Theory of Strongly Semantic Information (TSSI). Again, the goal is to develop a theory from which semantic information can be quantified, which would clearly be a useful theory for the engineering of semantic communication. As a starting point, three desiderata are given:
\begin{enumerate}
    \item[D.1] Avoid any counterintuitive inequality comparable to BCP
    \item[D.2] Treat the alethic (truth) of a sentence not as a supervenient but as a necessary feature
    \item[D.3] Extend a quantitative analysis to the whole family of information-related concepts
\end{enumerate}

The core of TSSI is the definition of degrees of vacuity and inaccuracy. The intuitive idea is that semantic information is related to ``how true'' and ``how false'' a sentence is. A highly vacuous sentence is one that is true, but carries with it little information. Similarly, a highly inaccurate sentence is false and also carries little information. This brings about the strange idea that a false sentence may carry more information than a true sentence. As an example, consider the system of three individuals and two predicates described earlier. The tautology $i$ = ``($a$ is $F$) or ($a$ is $M$)'' clearly provides no information, yet it is always true nonetheless. In contrast, consider the sentence $j$ = ``($a$ is $F$) and ($b$ is $M$) and ($c$ is $F$) and ($a$ is $Y$) and ($b$ is $Y$) and ($c$ is $O$)'' when $c$ is actually $M$. While $j$ is false, it is not too difficult to reason that it carries more information than $i$.

Mathematically, these concepts are formalized as a positive or negative degree of ``semantic distance'' of a sentence $i$ from a fixed point, which is defined as the given situation $w$ to which $i$ is supposed to refer. True statements take on positive degrees between 0 and 1, while false statements take on negative degrees between $-1$ and 0. This mapping is denoted by the function $f(i)$. For false statements, the \textit{degree of inaccuracy} simply counts the number of false atomic statements $e$ in $i$ and divides by the total number of atomic statements, or the \textit{length} $l$ of $i$
\begin{equation}
    \label{eq_deg_inac}
    f(i) = -e(i)/l(i),
\end{equation}
where $i$ is a false sentence. On the other hand, the \textit{degree of vacuity} is more difficult to define since all atomic statements of a true sentence are true. Therefore, this degree is defined as the number of situations $n$ (including the true situation) with which $i$ is consistent divided by the total number of possible situations ($s^l$ for a system with $s$ predicates),
\begin{equation}
    \label{eq_deg_vac}
    f(i) = n(i)/s^l,
\end{equation}
where $i$ is a true sentence, and a ``situation'' is nothing more than a complete state description as was defined for TWSI. Note the lack of symmetry between (\ref{eq_deg_inac}) and (\ref{eq_deg_vac}); this is one argument against TSSI.

\begin{figure}[htbp]
    \centering
    \includegraphics[scale = 0.50]{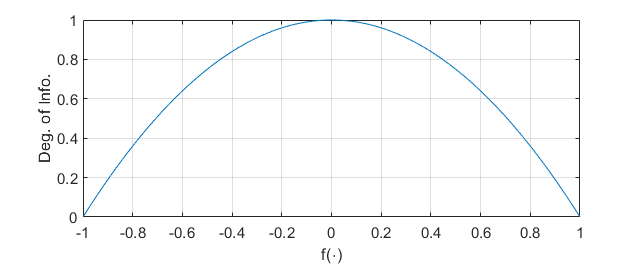}
    \caption{Relationship between degrees of inaccuracy, vacuity, and informativeness}
    \label{fig_tssi_deg_info}
\end{figure}

Using these degrees, the \textit{degree of informativeness} is defined as
\begin{equation}
    \imath(i) = 1 - f^2(i).
\end{equation}
Continuing the previous example, the tautology is consistent with all situations, and thus has a degree of vacuity $f(i) = 2^6/2^6 = 1$ and a degree of informativeness of $\imath(i) = 1 - 1^2 = 0$. Meanwhile the false statement $j$ has degree of inaccuracy $f(j) = -1/6$ and a degree of informativeness $\imath(j) = 1 - (-1/6)^2 \approx 0.972$, and we see that $\imath(\cdot)$ is consistent with intuition. The relationship between degrees of vacuity and inaccuracy and degree of informativeness is given in Figure \ref{fig_tssi_deg_info}.

The \textit{amount of vacuous information} is then defined as the integral of this curve from $0$ to the degree of vacuity; the maximum amount of vacuous information is then simply the integral from 0 to 1, yielding 2/3. Defining this maximum as $\alpha$ and the amount of vacuous information carried by $i$ as $\beta(i)$, the amount of strongly semantic information carried by $i$ is defined as
\begin{equation}
    \label{eq_flor_in}
    \imath^*(i) = \alpha - \beta(i)
\end{equation}

It is finally argued that $\imath^*(i)$ provides a solution to the BCP, in that it shows that ``semantic information about a situation presents an actual possibility that is inconsistent with at least one but not all other possibilities. A contradiction is not information-rich because it is not a possibility.'' It is stated that in TSSI, a contradiction is simply a limit instance of ``uninformation,'' or lack of both positive and negative misinformation. 

\subsection{Semantic Information with Truthlikeness}

TSSI is a step in the right direction, but still has some shortcomings with regards to quantifying semantic information. First, the degrees of vacuity and inaccuracy are inherently asymmetric; the former is a measure depending on the model, and the second is a measure based solely on the sentence at hand. In addition, it is unclear how to quantify vacuity and inaccuracy for more complex sentences beyond simple conjunctions of atomic statements.

In \cite{dalfonso}, Simon D'Alfonso builds on the foundations laid by TWSI and TSSI by expanding on Floridi's idea of using ``truthlikeness'' for quantifying semantic information. Two existing approaches to quantifying truthlikeness are offered. The first, termed the Tichie-Oddie approach, computes truthlikeness as the compliment of some distance function $\Delta(\cdot)$ from the statement $i$ and the true statement $t$ \cite{dalfonso}:
\begin{equation}
    \label{eq_to_tr}
    \Tr(i) = 1 - \Delta(i,t)
\end{equation}
where $\Delta(\cdot)$ takes values in $[0,1]$. We can see that (\ref{eq_flor_in}) is a particular case of (\ref{eq_to_tr}) with $\Delta(i,t) = f^2(i)$. Let $\mathcal{W}_i$ denote the set of states in which statement $i$ is true. Furthermore, define $\delta_i^{(t)}$ as the number of differing atomic statements in state $w_i$ and the true state $w_t$, i.e., the number of false atomic statements in $w_i$. This value is weighted by the inverse of the number of propositions in the universe. The Tichie-Oddie approach \cite{dalfonso} specifies the distance function as
\begin{equation}
    \label{eq_to_delta}
    \Delta(i,t) = \frac{\sum_{w_i \in \mathcal{W}_i}\delta_i^{(t)}}{\left\vert \mathcal{W}_i \right\vert}
\end{equation}

{\renewcommand{\arraystretch}{1.3}
\begin{table*}[htbp]
    \centering
    \caption{Summary of classical semantic information measures}
    \begin{tabular}{|p{.11\linewidth}|p{.18\linewidth}|p{.32\linewidth}|p{.29\linewidth}|}\hline
    \multirow{1}{\linewidth}{\centering \textbf{Method}}                          &  \multirow{1}{\linewidth}{\centering \textbf{Measure}}                                                                                  &  \multirow{1}{\linewidth}{\centering \textbf{Benefits}}                                                                       & \multirow{1}{\linewidth}{\centering \textbf{Drawbacks}} \\\hline
    \multirow{4}{*}{TWSI {\csname @fileswfalse\endcsname\cite{carnap}}}            & \multirow{2}{\linewidth}{ $\cont(i)$}                                                     & \tabitem  Simple measure satisfying (\ref{eq_inam_r1})-(\ref{eq_inam_r3})                 & \tabitem  Does not satisfy independent additivity\\
                                    &                                                                                           & \tabitem  Entirely dependent on logical probabilities                                     & \tabitem  Trouble dealing with contradictions (BCP)\\\cline{2-4}
                                    & \multirow{2}{\linewidth}{ $\inf(i) = \log_2 \frac{1}{1 - \cont(i)}$}                      & \tabitem  Satisfies (\ref{eq_inam_r1})-(\ref{eq_inam_r3}) and independent additivity  & \tabitem  Does not satisfy triangle inequality\\
                                    &                                                                                           & \tabitem  Entirely dependent on logical probabilities                                 & \tabitem  Trouble dealing with contradictions (BCP)\\\hline
    \multirow{4}{*}{TSSI {\csname @fileswfalse\endcsname\cite{floridi}}}           & \multirow{4}{\linewidth}{$f(i) = -e(i)/l(i)$ ($i$ false), $f(i) = n(i)/s^l$ ($i$ true)}   & \multirow{2}{\linewidth}{\tabitem  Avoids BCP by considering truth}                                             & \tabitem  Asymmetric: different measures for true and false statements\\
                                    &                                                                                           & \tabitem  Satisfies intuition that true and false statements carry information        & \tabitem  Unclear how to define for more complex statements \\\hline
    \multirow{6}{*}{Truthlikeness {\csname @fileswfalse\endcsname\cite{dalfonso}}}  & \multirow{3}{\linewidth}{$\Tr(i) = 1 - \frac{\sum_{w_i \in \mathcal{W}_i}\delta_i^{(t)}}{\left\vert \mathcal{W}_i \right\vert}$} & \tabitem  Avoids BCP by considering truth                                             & \multirow{3}{\linewidth}{\tabitem Addition of false statement can increase information yield}  \\
                                    &                                                                                           & \tabitem Provides flexibility through adjustment of atom weights          &  \\\cline{2-4}
                                    & \multirow{2}{\linewidth}{$\Tr(i) = 1 - \gamma \Delta_{\textrm{min}}(i,t) + \lambda \Delta_{\textrm{sum}}(i,t)$} & \tabitem  Avoids BCP by considering truth                                             & \multirow{2}{\linewidth}{\tabitem May not be applicable in certain scenarios}  \\
                                    &                                                                                           & \tabitem Satisfies several adequacy conditions           &  \\\hline
    \end{tabular}
    \label{tab_classic_measures}
\end{table*}
}

Continuing with the example in the previous subsection, $i$ agrees with all state descriptions, so $\vert \mathcal{W}_i \vert = 64$. After computing the sum in the numerator we have $\Delta(i,t) = 32/64 = 1/2$ regardless of the true state. Thus, a statement that is always true provides little information no matter the true state of the universe. For the second involving sentence $j$, we obtain $\Delta(j,t) = 1/6$ and $\Tr(j) = 5/6$, again matching intuition that the false $j$ carries more information than the true $i$.

In general, the function $\Delta(\cdot)$ can be any distance measure between the state specified by the sentence $i$ and that of the true statement $t$. The Niiniluoto approach \cite{dalfonso} makes use of a different distance metric, namely the min-sum measure
\begin{equation}
    \label{eq_niin_delta}
    \Delta_{\textrm{ms}}^{\gamma \lambda}(i,t) = \gamma \Delta_{\textrm{min}}(i,t) + \lambda \Delta_{\textrm{sum}}(i,t)
\end{equation}
where
\begin{align}
    \Delta_{\textrm{min}}(i,t) &= \min_{w_i \in \mathcal{W}_i} \Delta(w_i,w_t)\\
    \Delta_{\textrm{sum}}(i,t) &= \frac{\sum_{w_b \in \mathcal{W}_i}\Delta(w_i,w_t)}{\sum_{w_b \in \textbf{B}}\Delta(w_b,w_t)}
\end{align}
and $\textbf{B}$ is the set of all states in the logical space. The state distance is defined as
\begin{align}
    \Delta(w_i,w_t) = \frac{\delta_i^{(t)}}{n}
\end{align}
where $n$ is some atomic weight. Niiniluoto shows that the min-sum measure of (\ref{eq_niin_delta}) satisfies certain adequacy conditions.

Finally, D'Alfonso proposes a novel measure, termed the \textit{value aggregate} measure, which is claimed to lie in between the Tichy/Oddie and Niiniluoto approaches. First, each state is assigned a value, with
\begin{equation}
    \val(w) = \frac{t^{(w)}}{n2^n}
\end{equation}
where $t^{(w)}$ is the number of true atoms in state $w$, and $n$ is the number of propositional variables in the logical space. The following algorithm is used to calculate information yield for a statement $i$:
\begin{enumerate}
    \item[1.] Determine $\mathcal{W}_i$.
    \item[2.] Place members of  $\mathcal{W}_i$ into an array $X_1$ and order from lowest to highest value.
    \item[3.] Create empty array $X_2$ of length $2^n$. Fill the first $\vert \mathcal{W}_i \vert$ elements with the array $X_1$. Use the last (highest value) element of $X_1$ to fill the remain spaces of $X_2$.
    \item[4.] Sum the values of $X_2$ to get the information measure.
\end{enumerate}
It is claimed without proof that the value aggregate measure satisfies many of the adequacy conditions listed by Niiniluoto. A summary of the measures discussed in this section is provided in Table \ref{tab_classic_measures}.

\subsection{Extensions of TWSI}

These approaches to quantifying semantic information, namely TWSI, TSSI, and truthlikeness, all provide the groundwork for a theory of semantic information, which could enable semantic communication. In \cite{bao}, TWSI is used as a foundation for a general, abstract semantic communication model. First a semantic information source is defined as a tuple $(W_s, K_s, I_s, M_s)$, where
\begin{itemize}
    \item $W_s$ is the world model (potentially observable by the source),
    \item $K_s$ is the background knowledge of the source,
    \item $I_s$ is the inference procedure of the source,
    \item $M_s$ is the message generator used by the source.
\end{itemize}
Similarly, the semantic receiver is defined by the analogous tuple $(W_r, K_r, I_r, M_r)$. The source model is further specified by assuming that the world model $W_s$ is a set of interpretations with probability distributions $\mu$; for the familiar example of propositional logic, an interpretation would be a set of positive propositions. The inference procedure $I_s$ is then defined as a satisfiability reasoner for the propositional logic, and the message generator $M_s$ employs a fixed coding strategy. The model entropy is then given by
\begin{equation}
    \label{eq_model_ent}
    H(W) = -\sum_{w \in W} \mu(w) \log_2 \mu(w)
\end{equation}

Letting $m(x)$ denote the logical probability of a message $x$ within this model, the \textit{semantic entropy} of $x$ is defined as
\begin{equation}
    \label{eq_sem_ent}
    H_s(x) = -\log_2(m(x)).
\end{equation}
Note that this is equivalent to the TWSI measure given by (\ref{eq_carnap_inf}) with $m(x) = 1 - \cont(x)$. When the knowledge base $K_s$ is included in this formulation, the set of possible worlds is restricted to a set compatible with $K_s$. This brings about the notion of \textit{conditional semantic entropy}, where $m(x\vert K)$ now denotes the logical probability of $x$ conditioned upon the background knowledge, and we have
\begin{equation}
    \label{eq_cond_sem_ent}
    H_s(x \vert K) = -\log_2(m(x\vert K)).
\end{equation}

\begin{figure*}[htbp]
    \centering
    \includegraphics[scale = 0.35]{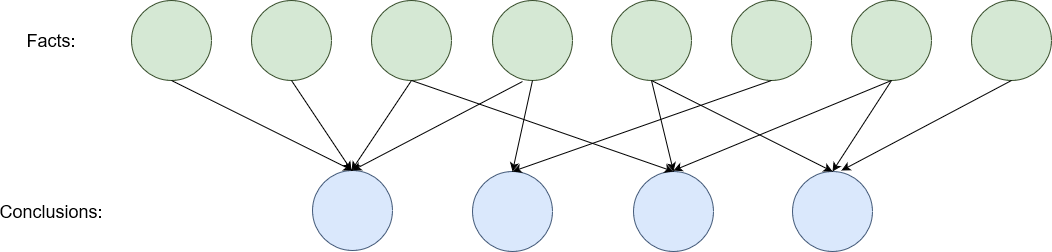}
    \caption{Bipartite fact-conclusion graph model of a knowledge base}
    \label{fig_fc_kg}
\end{figure*}

Letting $X$ be a finite set of allowed messages with probability distribution $P(X)$, we know the classic Shannon entropy of $X$ as
\begin{equation}
    \label{eq_shan_ent}
    H(X) = -\sum_{x\in X} P(x) \log_2 P(x).
\end{equation}
The following theorem relates the model (semantic) entropy (\ref{eq_model_ent}) to the message (syntactic) entropy (\ref{eq_shan_ent}):
\begin{thm}
\label{thm_sem_syn_ent}
$H(X) = H(W) + H(X\vert W) - H(W \vert X)$.
\end{thm}
\begin{proof}
See \cite{bao}.
\end{proof}
The main implication of Theorem \ref{thm_sem_syn_ent} is a formal method of quantifying semantic uncertainty that is rooted in the TWSI and relating it to classic Shannon entropy. As entropy is used to quantify information in classic Information Theory, this gives a way of comparing the syntactic information and the semantic information under the present world model.

The authors in \cite{bao} then discuss the idea of using semantics for data compression. Intuitively, the idea is that some messages may be semantically equivalent without being syntactically equivalent. For example, many times we are able to understand the true meaning of text despite some minor errors in spelling or grammar. Hence, a syntactic error does not necessarily induce a semantic error. Based on this principle, the idea is that we can achieve maximum compression by choosing the smallest semantically equivalent message for communication. Let $\overline{X}$ denote the smallest subset of $X$ such that each $x \in X$ is semantically equivalent to some $\overline{x} \in \overline{X}$.
\begin{thm}
\label{thm_sem_source_code}
For a semantic source with interface language $X$, there exists a coding strategy to generate a semantically equivalent interface language $X'$ with message entropy $H(X') \geq H(\overline{X})$. No such $X'$ exists with message entropy $H(X) < H(\overline{X})$.
\end{thm}
\begin{proof}
See \cite{bao}.
\end{proof}
Theorem \ref{thm_sem_source_code} provides bounds on the maximum achievable data compression give a model as described above. Note that Theorem \ref{thm_sem_source_code} is analogous to the classical source coding theorem, in that it gives existence of such a code but no insight into how to design the coding strategy.

The final major result of \cite{bao} is the so-called Semantic Channel Coding Theorem. Some notations used include:
\begin{itemize}
    \item $I(X;Y) = H(X) - H(X\vert Y)$ is the traditional information-theoretic \textit{mutual information} between $X$ and $Y$
    \item $\overline{H_s(Y)} = -\sum_y p(y) H_s(y)$ is the average logical information of received messages 
\end{itemize}
\begin{thm}
\label{thm_sem_chan_code}
For every discrete memoryless channel, the channel capacity
\begin{equation}
    C_s = \sup_{P(X\mid W)} \{ I(X; Y) - H(W \vert X) + \overline{H_s(Y)} \}
\end{equation}
has the following property: For any $\epsilon > 0$ and $R < C_s$, there is a block coding strategy such that the maximal probability of semantic error is $< \epsilon$.
\end{thm}
\begin{proof}
See \cite{bao}.
\end{proof}

Similar to Theorem \ref{thm_sem_source_code}, Theorem \ref{thm_sem_chan_code} provides a result that parallels the classic Channel Coding Theorem of Information Theory. This result gives a bound on the maximum amount of information that can be transmitted for some arbitrary probability of semantic error. 

In \cite{basu}, the general model and results of \cite{bao} are extended and practical semantic compression algorithms are given based on graph theoretic results. However, basic definitions, such as the semantic source and receiver, are different from those in \cite{bao}, making it difficult to relate this work to previous results. Perhaps the most useful contribution of \cite{basu} is a discussion of some practical techniques for semantic compression.

The first suggested idea is to allow \textit{non-uniquely decodable} codes. This fits the case in which a certain syntactic message represents two semantically equivalent states, i.e., message $x$ could be decoded as either state $a$ or $b$. Another idea is to extend the concept of erasure channel codes, such that some bits are intentionally ``erased.'' With this approach, only partial information may be recovered at the receiver, with some intentional semantic ambiguity.

A practical algorithm is then suggested for a system in which the source and the receiver share a knowledge base that is defined as a bipartite fact-conclusion graph, see Figure \ref{fig_fc_kg}. The problem studied is that in which the source wants to convey a set of conclusions to the receiver in as few symbols as possible (where both facts and conclusions can be transmitted and are equally expensive). This problem reduces to computing the \textit{minimum-vertex cover} which is solvable in polynomial time for bipartite graphs. This simple example represents a situation in which semantic compression is computationally feasible, given the assumption on background knowledge.

\subsection{Summary}
The classical semantic information approach attempts to mirror the path followed in the monumental development of Information Theory to define and quantify semantic information. TWSI attempts to achieve this through the use of logical probabilities rather than statistical probabilities. However, this theory contains a paradox in which a contradiction carries maximal semantic information. As a remedy, TSSI introduces the use of truth values to quantify semantic information, based on the idea that both false and true statements can carry varying degrees of information. This idea has been extended through the introduction of various truthlikeness measures for quantifying semantic information. Recently, some have attempted to extend the original TWSI, relating semantic entropy to traditional information-theoretic entropy and providing analogous source-coding and channel-coding theorems. All of these approaches seek to quantify semantic information of a sentence through the use of logical probabilities and truth values.

With regards to the motivating problem described in Section \ref{sec_intro}, namely, the trend of exponentially increasing global data traffic, the works described in this section provide no quantitative results addressing this issue. This is due to the fact that these works are more concerned with the development of semantic information theory, rather than the subsequent use of this theory for semantic communication. That is not to say that the presumed benefits are nonexistent, however, and future work in this area should seek to confirm this potential.

\subsection{Challenges and Opportunities}
\label{subsec_classic_c_and_os}

Given the ability to quantify semantic information, development of optimal coding techniques (source and channel) should follow. However, as can be seen from the previous discussion, this quantification is no trivial task. A definition of semantic information itself is elusive, and attempts to quantify it become mired with paradoxes and counter-intuitive results. However, should such a complete theory exist, the benefits that would follow are clear, thus presenting a major opportunity.

Previous work seems to point to the idea of truthlikeness as the best approach to quantifying semantic information. The main challenge with this approach is the need for a large knowledge base. Simple examples are given using propositional logic models consisting a few objects and propositions, but the size of these models explodes for more realistic and practical systems. Furthermore, these models must be predefined and known at both the source and receiver, which introduces further complications. Another challenge is selection of the correct information measure. As we have seen, there is no single measure for semantic information (yet); the available measures all come with their respective strengths and shortcomings.

{\renewcommand{\arraystretch}{1.5}
\begin{table*}[t]
    \centering
    \caption{Summary of Works in Knowledge Graph-Based Semantic Communication}
    \begin{tabular}{|c|c|c|}\hline
    \multirow{3}{*}{Knowledge Representation}   & Delgado-Frias \& Moore {\csname @fileswfalse\endcsname\cite{delgadofrias}}  & ``Semantic network'' as a knowledge graph \\\cline{2-3} 
                                                & Mertoguno \& Lin {\csname @fileswfalse\endcsname\cite{mertoguno}} & Evolution of a distributed knowledge base \\\cline{2-3} 
                                                & Swartout \etal {\csname @fileswfalse\endcsname\cite{Swartout}} & Using ontologies to share knowledge \\\hline 
    \multirow{4}{*}{Semantic Similarity}        & Rada \etal {\csname @fileswfalse\endcsname\cite{rada}} & Link-based similarity measure on knowledge graphs \\\cline{2-3} 
                                                & Resnik {\csname @fileswfalse\endcsname\cite{Resnick}} & Node-based similarity measure on knowledge graphs \\\cline{2-3} 
                                                & Jiang \& Conrath {\csname @fileswfalse\endcsname\cite{jiang_semsim}} & Similarity measure considering both links and nodes \\\cline{2-3} 
                                                & Sathya \& Uthayan {\csname @fileswfalse\endcsname\cite{sathya_semsim}} & Measure of quality for an entire ontology \\\hline 
    \multirow{6}{*}{Semantic Sensor Web}    & Sheth \etal {\csname @fileswfalse\endcsname\cite{sheth_sensorweb}} & Semantic sensor web \\\cline{2-3} 
                                                & Gyrard \etal {\csname @fileswfalse\endcsname\cite{gyrard_M2M}} & Semantic sensor web for machine-to-machine communication \\\cline{2-3} 
                                                & Chun \etal {\csname @fileswfalse\endcsname\cite{chun_sem_iot}} & IoT directory for the semantic sensor web \\\cline{2-3} 
                                                & Bhajantri \& Pundalik {\csname @fileswfalse\endcsname\cite{bhajantri_dataproc}} & Data-processing for the semantic sensor web \\\cline{2-3} 
                                                & Schachinger \& Kastner {\csname @fileswfalse\endcsname\cite{schachinger_semM2M}} & Semantic interface for building automation \\\cline{2-3} 
                                                & Lakka \etal {\csname @fileswfalse\endcsname\cite{lakka_E2E}} & Interoperability of semantic systems \\\hline
   \multirow{7}{*}{Semantic Communication}      & Jeong \etal {\csname @fileswfalse\endcsname\cite{jeong_speech}} & Semantic error correction of spoken queries \\\cline{2-3} 
                                                & Zhang \etal {\csname @fileswfalse\endcsname\cite{zhang_ernie}} & Model for semantic natural language processing  \\\cline{2-3} 
                                                & Li {\csname @fileswfalse\endcsname\cite{li_dickinsons}} & Text analysis and character recognition with a KG and ML \\\cline{2-3} 
                                                & Wang \etal {\csname @fileswfalse\endcsname\cite{wang_KG_recommender}} & Hybrid KG/ML model for explanation of recommendations  \\\cline{2-3} 
                                                & Aumayr \etal {\csname @fileswfalse\endcsname\cite{aumayr_KG_recommender}} & Recommendation system for wireless network operation \\\cline{2-3} 
                                                & He \etal {\csname @fileswfalse\endcsname\cite{he_KG_agents}} & Model for general goal-based communication \\\cline{2-3} 
                                                & G\"uler \etal {\csname @fileswfalse\endcsname\cite{guler_sem_comm_game}} & Optimal semantic communication with game theory \\\hline 
    \multirow{3}{*}{Working with Knowledge Graphs}  & Wei \etal {\csname @fileswfalse\endcsname\cite{wei_KG_reasoning}} & Reasoning over large knowledge graphs \\\cline{2-3} 
                                                    & Zheng \etal {\csname @fileswfalse\endcsname\cite{zheng_KG_embed}} & Embedding of large-scale knowledge graphs \\\cline{2-3} 
                                                    & Zhu \etal {\csname @fileswfalse\endcsname\cite{zhu_KF_fuse}} & Fusion of knowledge graphs \\\hline 
    \end{tabular}
    \label{tab_kg_sec_summary}
\end{table*}
}

A clear opportunity lies in the further development of the theory of semantic information. While significant progress has been made on the subject, a unifying and ubiquitous theory does not yet exist. Another promising opportunity is the incorporation of existing semantic information measures into practical systems. In particular, \textit{neurosymbolic AI} has become a field of great interest as of late \cite{garcez_neurosymbolic_2020}. By combining the strengths of symbolic logic and DL, neurosymbolic AI could enable powerful learning systems that are capable of logical reasoning. By applying neurosymbolic AI to the problem of semantic communication, perhaps the logical model at the source and receiver could be learned from data. Then, the DL architecture could be used to efficiently facilitate semantic communication. As all of the previous work on semantic information quantification relies to some degree on propositional logic, the application of neurosymbolic AI seems to be a perfect fit.

\section{Knowledge Graph-Based Semantic Communication}
\label{subsec_ontol}

Semantic communication necessarily requires a knowledge base at both the transmitter and receiver. The engineering of semantic communication thus requires some form of knowledge representation to encapsulate the knowledge at the transmitter and receiver. One prevalent method of representing knowledge is through the use of a \textit{knowledge graph} (KG). Although a particular KG may be defined in different ways, in general a KG can be said to use a graph structure to model a given knowledge base. For example, nodes of the graph could represent objects, while edges represent relations between the objects. By referencing the KG, a system can then perform communication that is semantically-aware. The works discussed in this section are summarized in Table \ref{tab_kg_sec_summary}.

\subsection{Knowledge Representation with KG's}

\begin{figure}[htbp]
    \centering
    \includegraphics[scale = 0.30]{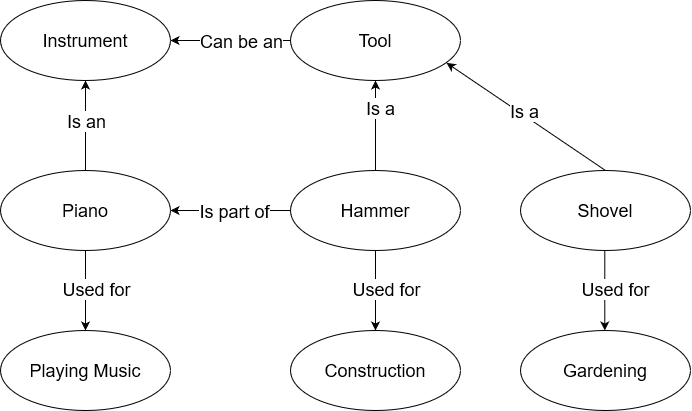}
    \caption{An example knowledge graph}
    \label{fig_kg}
\end{figure}

One of the earliest works making use of this idea was published in 1989, which proposes a so-called ``semantic network architecture'' for AI processing \cite{delgadofrias}. Arguing that knowledge representation and manipulation is required for artificial intelligence, a multi-processor architecture is proposed. This semantic network takes the form of a KG, in which nodes are defined as \textit{concepts} and edges are defined as \textit{relationships} between concepts. Figure \ref{fig_kg} gives a simple example of a KG. The architecture is composed of a grid of processing elements (PE's) which compute the corresponding links and nodes associated with the semantic network. By building the architecture in such a manner, it is argued that the system can perform intelligent actions, such as labelling the scene of a given image (computer vision). 

Other early studies into knowledge representation with KG's include \cite{mertoguno} and \cite{Swartout}, published in 1996 and 1997, respectively. In \cite{mertoguno}, the problem of a \textit{distributed} knowledge base is studied using KG's. Specifically, the challenges of evolving the distributed knowledge and controlling this evolution are considered. This distributed knowledge base is framed as a multi-agent system in which some communication exists between the agents. Within the system, each agent possesses a simple KG and agents collaborate to perform global inferencing. Participating agents receive distributed rewards after correct inferences, resembling a mutli-agent reinforcement learning scheme. A crossover operator is adopted from genetic algorithms to facilitate knowledge evolution. Distributed knowledge bases are also the focus of \cite{Swartout}, which uses the idea of a large-scale shared ontology. Here, an ontology is defined as ``a hierarchically structured set of terms for describing a domain,'' from which a knowledge base can be constructed. The key idea is that if two knowledge bases are formed from the same ontology, knowledge can be easily shared between the two. An analogy would be two people from the same culture; the shared cultural norms and ideas would serve as the so-called ontology, and though their knowledge bases will not be the exact same, intuitively it will be easier for them to communicate given their shared background. A set of desiderata for ontologies is given, essentially proposing that ontologies should represent large-scale, living documents from which we can define smaller, application-specific knowledge bases. This idea is the basic premise behind the Web Ontology Language (OWL) standards that shape the semantic web \cite{owl}.

\subsection{Semantic Similarity Measures on KG's}

As was seen in the previous section, an important idea for semantic communication is the idea of semantic similarity. In the sense of classical semantic information, logical probabilities and truthlikeness were used to define this similarity. How should one quantify semantic similarity given a KG representation? This is a question that some began to address around the time KG's emerged as a way of representing knowledge. In \cite{rada}, a metric of semantic similarity, or distance, is proposed and simply termed Distance. The authors assume a KG that consists of an \textit{is-a} hierarchy, see Figure \ref{fig_is_a_kg} below.
\begin{figure}[htbp]
    \centering
    \includegraphics[scale = 0.40]{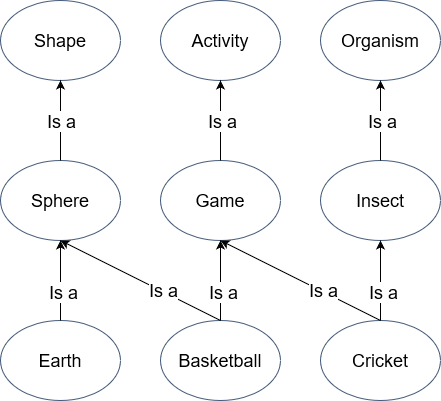}
    \caption{Example of an \textit{is-a} KG, where all edges represent an \textit{is-a} relationship}
    \label{fig_is_a_kg}
\end{figure}
This Distance metric is defined for \textit{sets} of concepts and is dependent upon the path lengths between nodes in a KG; formally, it is defined as the average minimum path length over all pairwise combinations of nodes between two sets of nodes, i.e., for sets of nodes $\mathcal{X}$ and $\mathcal{Y}$,
\begin{equation}
    \label{eq_Distance}
    \Distance(\mathcal{X},\mathcal{Y}) = \frac{1}{\vert \mathcal{X} \vert \vert \mathcal{Y} \vert} \sum_{y \in \mathcal{Y}} \sum_{x \in \mathcal{X}} d( x,y),
\end{equation}
where $d(x,y)$ is the shortest path between $x$ and $y$. For example, consider the distance between the sets \{Sphere, Earth\} and \{Basketball\} in the KG shown in Figure \ref{fig_is_a_kg}. The shortest path between Sphere and Basketball is 1, while the shortest path between Earth and Basketball is 2, yielding a total distance of 3/2. Repeating for the sets \{Sphere, Earth\} and \{Cricket\}, we obtain a distance of 7/2, and thus we can conclude that the concepts \{Sphere, Earth\} are more similar to the concept \{Basketball\} than \{Cricket\}. It is shown through experimental results that (1) Distance can simulate human assessments of conceptual distance and (2) Distance can evaluate some cognitive aspects of semantic networks. However, it is also found that Distance is less applicable to nonhierarchical KG's.

An alternate measure is presented in \cite{Resnick}, again for an is-a taxonomy KG. First, it is noted that link-based measures, such as (\ref{eq_Distance}), suffer from the fact that links in the taxonomy are assumed to represent uniform distances, while in reality some linked concepts may be ``closer'' than others. Indeed, for the average person in the United States, \{Basketball\} probably shares a stronger intuitive link to \{Game\} than \{Cricket\} does, while this association may differ elsewhere. Therefore, a \textit{node-based} measure is proposed based on the notion of information content. First, the KG is augmented with a function $p: \mathcal{C} \rightarrow [0,1]$ where $\mathcal{C}$ is the set of all nodes in the graph. $p(c)$ can be thought of as the \textit{probability} of encountering a concept $c$; thus, concepts higher in the taxonomy will have greater probability. Then we can define the semantic similarity between two concepts as
\begin{equation}
    \label{eq_resnick_sim}
    \semsim(c_1, c_2) = \max_{c \in S(c_1, c_2)} -\log p(c)
\end{equation}
where $S(c_1,c_2)$ is the set of concepts that subsume \textit{both} $c_1$ and $c_2$. Intuitively, this measure computes the log-inverse probability of the \textit{most-specific} node (farthest ``down'' in the taxonomy) which branches to both concepts; therefore, the farther down this node, the smaller it's probability, and the greater the similarity measure. For example, in Figure \ref{fig_is_a_kg}, this would imply that \{Cricket\} is more similar to \{Basketball\} than to \{Earth\}.

{\renewcommand{\arraystretch}{1.3}
\begin{table*}[htbp]
    \centering
    \caption{Summary of KG-based semantic similarity measures}
    \begin{tabular}{|p{.15\linewidth}|p{.20\linewidth}|p{.27\linewidth}|p{.27\linewidth}|}\hline
    \multirow{1}{\linewidth}{\centering \textbf{Method}}                          &  \multirow{1}{\linewidth}{\centering \textbf{Measure}}                                                                                  &  \multirow{1}{\linewidth}{\centering \textbf{Benefits}}                                                                       & \multirow{1}{\linewidth}{\centering \textbf{Drawbacks}} \\\hline
    \multirow{2}{*}{Link-Based {\csname @fileswfalse\endcsname\cite{rada}}}            & \multirow{2}{\linewidth}{ $\Distance(\mathcal{X},\mathcal{Y}) = \frac{1}{\vert \mathcal{X} \vert \vert \mathcal{Y} \vert} \sum_{y \in \mathcal{Y}} \sum_{x \in \mathcal{X}} d( x,y)$}                                                     & \tabitem Simplest to compute                   & \tabitem Each link assumed as equi-distant  \\
                                    &                                                                                           & \tabitem Intuitive                                       & \tabitem No node-based information  \\\hline
    \multirow{2}{*}{Node-Based {\csname @fileswfalse\endcsname\cite{Resnick}}}           & \multirow{2}{\linewidth}{$\semsim(c_1, c_2) = \max_{c \in S(c_1, c_2)} -\log p(c)$}   & \tabitem Links not necessarily equi-distant                                               &  \multirow{2}{\linewidth}{\tabitem Requires additional function $p$} \\
                                    &                                                                                           & \tabitem Interpretable, based on probabilities         & \\\hline
    \multirow{2}{*}{Node/Link-Based {\csname @fileswfalse\endcsname\cite{jiang_semsim}}}  & \multirow{2}{\linewidth}{$\Dist(c_1,c_2) = \min_{\textrm{path}(c_1,c_2)} \sum_{c } \wt(c,p_c)$} & \tabitem Incorporates both ideas                                              & \tabitem Requires additional function $p$  \\
                                    &                                                                                           & \tabitem Achieves highest human correlation          &  \tabitem Complicated to compute \\\hline
    \end{tabular}
    \label{tab_kg_measures}
\end{table*}
}

Combining (\ref{eq_Distance}) and (\ref{eq_resnick_sim}), \cite{jiang_semsim} develops a new measure for the is-a taxonomy KG that provides an even higher correlation with human similarity judgement benchmarks. It involves a link-based calculation which takes into consideration node-based edge weights. First, information content is defined in the same way as \cite{Resnick}:
\begin{equation}
    \label{eq_IC}
    \IC(c) = -\log p(c).
\end{equation}
Then it is argued that the \textit{strength} of a child link is dependent on the information content of the parent node,
\begin{equation}
    \label{eq_LS}
    \LS(c,p_c) = \IC(c) - \IC(p_c),
\end{equation}
where $p_c$ is the parent node of $c$. This link strength, among other factors, is used to compute an overall edge weight $\wt(p_c,c)$ between the child $c$ and parent $p_c$. Finally, the semantic similarity, denoted as $\Dist$, is defined as summation of edge weights along the shortest path linking two nodes
\begin{equation}
    \label{eq_Dist}
    \Dist(c_1,c_2) = \min_{\textrm{path}(c_1,c_2)} \sum_{c \in \{ \textrm{path}(c_1,c_2) - \textrm{LSuper}(c_1,c_2)\} } \wt(c,p_c)
\end{equation}
where $\textrm{LSuper}(c_1,c_2)$ is the lowest super-ordinate of $c_1$ and $c_2$. Note that in \cite{jiang_semsim}, $\Dist(w_1,w_2)$ is used, where $w_1$ and $w_2$ are introduced to address the scenario when one node belongs to multiple inheritances. An evaluation of this metric shows a correlation value of 0.828 with human similarity judgements, higher than the node-based and link-based measures alone. This implies that considering link strength within a KG can lead to much more reasonable similarity judgements. A summary of the KG-based semantic similarity metrics described here are given in Table \ref{tab_kg_measures}.

While (\ref{eq_Distance}), (\ref{eq_resnick_sim}), and (\ref{eq_Dist}) all give ways to measure semantic similarity within a KG, a more recent work \cite{sathya_semsim} proposes a semantic metric to assess the quality of an entire ontology. It is argued that most existing metrics for ontology assessment consider only structural properties, and ignore the semantics of the ontology. The proposed metric is termed the ``relationship deviation metric'' and is determined by the number of breadthwise and depthwise relationships in the KG. It is shown that the proposed metric captures the quality of some example ontologies.

\subsection{The Semantic Sensor Web}

One prevalent use of KG's for semantics is the so-called semantic sensor web (SSW) proposed  in \cite{sheth_sensorweb}, which uses ideas from the semantic web. Namely, metadata is captured along with the desired data from each sensor, such that better sense can be made of the observations (similar to how metadata can be used to determine what is contained within a webpage). For example, providing the total lifetime of a sensor may lead to more informed decision-making; a sensor that has been operating well past its expected lifetime may be more likely to produce faulty measurements. With regards to communication, this metadata can provide the \textit{context} necessary for semantically efficient communication within the network. The suggested core set of attributes in \cite{sheth_sensorweb}, as adopted from the RDFa language \cite{rfda_core}, are
\begin{itemize}
    \item \textit{about}: a triple that specifies the resource metadata is about
    \item \textit{rel} and \textit{rev}: specify a relationship or reverse-relationship with another resource
    \item \textit{href}, \textit{src}, and \textit{resource}: specifies the partner resource
    \item \textit{property}: specifies a property for the content of an element
    \item \textit{instanceof}: optional, specifies the RDF type of the object
\end{itemize}
Furthermore, \cite{sheth_sensorweb} advocates for the use of ontologies along the three types of semantics associated with sensor data: spatial, temporal, and thematic. Once these ontologies have been defined, rule-based reasoning can be implemented to provide better inferences from sensor observations.

Building on this idea of an SSW, \cite{gyrard_M2M} proposes a semantic-based approach to automatically combine, enrich, and reason about machine-to-machine (M2M) data to support IoT applications. One key idea is that the \textit{meaning} of new information is pre-defined in an ontology, and therefore the ontology can facilitate the fusing of cross-domain knowledge. Suppose we possess an ontology which stores weather-related knowledge and soil-related knowledge and the relations between the two; by leveraging this shared knowledge, it might be possible to achieve smarter fusion of the data for agricultural decision-making. A concept called ``Linked Open Rules'' is defined as a means of sharing and reusing semantic rules, and some examples are given that demonstrate the proposed concept, including a weather monitoring application. In \cite{chun_sem_iot}, an IoT directory, called IoT-DS, is proposed to support semantic description, discovery, and integration of new objects as an alternative approach to building a SSW. One key difference is that IoT-DS distinguishes static and dynamic components, based on whether other attributes vary with time. It is shown that IoT-DS provides a 40\% reduction in communication overhead as compared to a naive approach.

\begin{figure*}[htbp]
    \centering
    \includegraphics[scale = 0.3]{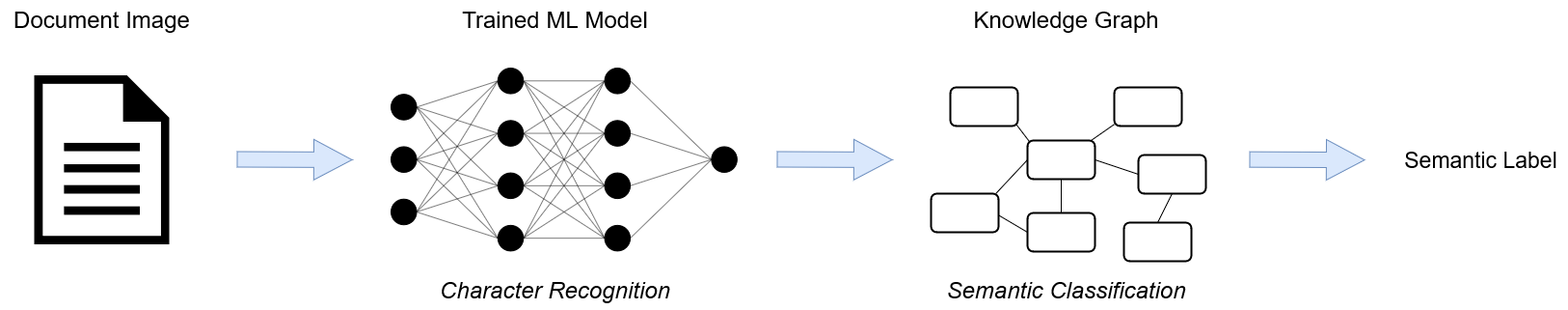}
    \caption{Simplified diagram of the framework proposed in \cite{li_dickinsons} for semantic document analysis}
    \label{fig_kg_semlabel}
\end{figure*}

Some more recent works that look at the idea of a SSW are \cite{bhajantri_dataproc, schachinger_semM2M, lakka_E2E}. In \cite{bhajantri_dataproc}, a survey of data processing techniques for SSW is provided, as well as another take on the architecture of a SSW, which is partitioned into physical, semantic, application and controller layers. \cite{schachinger_semM2M} considers again the problem of a semantic interface for M2M communication, this time with the intended application of building automation. Requirements are defined for the building automation problem, from which an interface is developed and an ontology formed. \cite{schachinger_semM2M} provides a great example of how KG-based semantics can be utilized to engineer a SSW for a specific application. Finally, \cite{lakka_E2E} studies the general problem of semantic interoperability, or the ability to interact and exchange data with shared meaning, between systems. An interoperability mechanism termed SEMIoTICS is proposed, in which an IoT application request is received by a directory which then connects the corresponding sensor and actuator to fufill the request.

\subsection{Knowledge Graphs for Semantic Communication}

The works discussed in the previous subsection are similar to the idea of the semantic web, in that they focus more on how to semantically describe objects rather than the task of semantic communication itself. One of the first works utilizing a KG for semantic communication is \cite{jeong_speech}, which focuses on semantic error correction for spoken query processing. Spoken query processing, or question answering, has become a hot topic as of late (more on this in Section \ref{subsec_ml}). \cite{jeong_speech} proposes the use of two KG's: a \textit{domain} dictionary and an \textit{ontology} dictionary. The first represents application-specific knowledge, while the second contains the pure general knowledge of the world. In the agricultural setting mentioned above, the ontology dictionary might contain general information that does not vary between crops or location, while the domain dictionary might consist of site-specific information. For a spoken query, the semantic recovery stage involves the use of a semantic confusion table based on the domain knowledge to replace semantic errors. Lexical, or syntactic, recovery is then performed based on the corrected semantic phrase. Experiments performed on the domain of in-vehicle telematics show that the technique yields a 37\% reduction in term-error-rate as compared to baseline models. This decreased error rate bodes well for traffic reduction in communication systems, as less information will need to be retransmitted due to errors.

In recent years, some have sought to apply KG-based semantics to text-based communication. \cite{zhang_ernie} looks to apply KG's to the problem of natural language processing (NLP), and proposes an enhanced language representation model termed ERNIE, which is an enhancement of the popular NLP model BERT \cite{devlin-etal-2019-bert}. The model operates by first recognizing entities in some text, and matching these to entities in a pre-defined KG. The KG representation is then embedded using known algorithms, such as TransE \cite{2013_bordes_transE}, and then used in conjunction with standard text embeddings as the input to an aggregator. DL encoder-decoder techniques are then used to perform common NLP tasks. In \cite{li_dickinsons}, a hybrid KG-ML approach is proposed to perform text analysis through character recognition. The model first uses DL to perform character recognition over two bodies of text, then uses semantic measures to quantify how similar the bodies of text are. To quantify the semantic similarity, each word is modeled as a word in a graph, and corresponding distance metrics are proposed. The general framework is illustrated in Figure \ref{fig_kg_semlabel}. Through experiments carried out on Dickinson's Portfolio, it is concluded that the performance of the proposed technique can meet real-time recognition requirements.

Another application of KG's for semantic communication involves recommendation systems, in which the goal is for an automated system to make the best possible recommendations to some user. One way that KG's have been used for this task is to enhance explainability of the recommendations \cite{wang_KG_recommender}. Explainability can enhance a user's experience when receiving recommendations, e.g., Amazon suggesting products to a user ``based on previous purchases.'' In \cite{wang_KG_recommender}, a model termed \textit{Knowledge Path Recurrent Network} (KPRN) is proposed as a hybrid KG-ML model. This model works on a KG which contains objects in the recommendation system and performs reasoning based on the paths in the KG to infer user preference. In the Amazon example, products would be modeled as nodes in the KG, and the software might suggest to a user a product with the shortest path length to that just purchased by the user. A long short-term memory (LSTM) network is adopted to model the sequential dependencies of objects and relations, from which a pooling operation is used to obtain the prediction. By modeling the sequential dependencies, the system can offer an explanation for each prediction. One scenario where a recommendation system is useful is wireless network management, where an automated system can recommend the best course of action to an operator. The growing complexities of these networks can make management difficult for a human operator, while automated methods might not always make the best decisions. Thus, a hybrid approach has some benefits. \cite{aumayr_KG_recommender} focuses on gathering context from a wireless network and correlating it with useful information from documents in the network provider's domain using KG's. The KG is formed from two types of documents: product troubleshooting manuals and incident/troubleshooting reports by technicians. The first provides well-structured problem solving instructions, while the second provides important links between symptoms and issues. This KG is then used to reason about new problems that arise, and recommend a course of action for the operator. Experiments show a decrease of up to 91\% of the documents that are presented to an operator, drastically reducing the amount of information involved in the communication process.

Finally, there has been some work on the more general scenario where two agents strategically communicate to achieve some goal, in which KG's are used to facilitate communication. \cite{he_KG_agents} studies a symmetric collaborative dialogue setting in which two agents communicate to achieve a common goal. Each agent possesses private knowledge. The model, termed Dynamic Knowledge Graph Network (DynoNet) models the dialogue state as a KG which evolves as the conversation advances. The graph contains three types of nodes, namely item, attribute, and entity nodes. These nodes are embedded, and used as an input to a LSTM network. Provided this embedding and an embedded utterance from another agent, the network generates an utterance in response. Experiments show that DynoNet is able to hold a coherent and strategic conversation with a human, and that the number of entities and attributes uttered to achieve the goal are reduced by 47\% and 17\%, respectively, when compared to baseline rule-based communication strategies. Goal-oriented communication naturally lends itself to game-theoretic analysis, which is the focus of \cite{guler_sem_comm_game}. In goal-oriented communication, as in game theory, there are two or more agents that seek to achieve some goal, and can employ multiple strategies to reach said goals. Similar to \cite{he_KG_agents}, in \cite{guler_sem_comm_game} communication is modeled as taking place between two agents, with the addition of a third agent who could aim to either improve/deteriorate communication performance. The optimal transmission policies are characterized, where optimality is defined as minimizing end-to-end semantic error. This error is derived from the semantic similarity measures proposed in \cite{rada,Resnick,jiang_semsim}. The interaction is modeled as a Bayesian game, where uncertainty is introduced about the characteristics of other agents. It is shown that, in the static scenario, that finding encoding/decoding strategies to minimize average semantic error is an NP-hard problem, and two algorithms are proposed. In addition, it is demonstrated that when the third agents signals its true nature to the communicating agents, a sequential equilibrium is attainable, i.e., when sufficient information is available regarding the intentions of agents involved in the communication, efficient semantic communication can be achieved. It is shown that judicious transmission policies can greatly reduce semantic errors.

\subsection{Working with KG's}

In the previous subsections, we have seen how KG's can be used to facilitate semantic communication. If our goal is to engineer semantic communication through the use of KG's, then we must develop effective methods of working with KG's. Reasoning over, i.e., deriving knowledge from, a KG is a challenge that becomes more difficult with increasing scale of the KG. In \cite{wei_KG_reasoning}, a reasoning system is proposed for large-scale KG's. This system, termed KGRL, is based on the web ontology language 2 rules logic (OWL2 RL) which was developed for the semantic web. Using the rules defined by OWL2 RL, the iterations of the reasoning procedure are reduced based on dependency relations and multiple applications of these rules. Experiments show that KGRL is able to greatly increase reasoning efficiency as compared to state-of-the-art reasoning systems.

{\renewcommand{\arraystretch}{1.5}
\begin{table*}[t]
    \centering
    \caption{Summary of Works in Machine Learning-Based Semantic Communcation}
    \begin{tabular}{|c|c|c|}\hline
    \multirow{11}{*}{Deep Learning}     & Lu \etal {\csname @fileswfalse\endcsname\cite{lu_radiotelephony}} & Ensuring mutual understanding using a LSTM-RNN \\\cline{2-3} 
                                        & Hua \& Du {\csname @fileswfalse\endcsname\cite{hua_deep_sem_corr}} & GAN for cross-modal retrieval \\\cline{2-3}
                                        & Huang \etal {\csname @fileswfalse\endcsname\cite{huang_dl_image_coding}} & Semantic coding of images with a GAN \\\cline{2-3}
                                        & Qiao \etal {\csname @fileswfalse\endcsname\cite{qiao_seed}} & CNN, RNN for scene text recognition \\\cline{2-3}
                                        & Tong \etal {\csname @fileswfalse\endcsname\cite{tong_audio_semcomm}} & Audio-based semantic communication with CNN and federated learning \\\cline{2-3}
                                        & Zhou \etal {\csname @fileswfalse\endcsname\cite{zhou_universal_tx}} & Text-based semantic communication using a transformer \\\cline{2-3}
                                        & Sana \& Strinati {\csname @fileswfalse\endcsname\cite{sana_semcomm_6G}} & End-to-end communication with semantic symbols using a transformer \\\cline{2-3}
                                        & Xie \etal {\csname @fileswfalse\endcsname\cite{xie_deepsc}} & Text-based semantic communication with DeepSC \\\cline{2-3}
                                        & Weng \etal {\csname @fileswfalse\endcsname\cite{xie_deepsc_speech}} & Speech-based semantic communication with DeepSC-S \\\cline{2-3}
                                        & Xie \etal {\csname @fileswfalse\endcsname\cite{xie_deepsc_vqa}} & Visual question answering with MU-DeepSC \\\cline{2-3}
                                        & Xie \& Qin {\csname @fileswfalse\endcsname\cite{xie_lite_deepsc}} & Semantic communication for IoT with L-DeepSC \\\hline 
   \multirow{4}{*}{Reinforcement Learning}      & Lu \etal {\csname @fileswfalse\endcsname\cite{lu_rl_semcomm}} & General semantic communication with arbitrary similarity function \\\cline{2-3} 
                                                & Wang \etal {\csname @fileswfalse\endcsname\cite{wang_kg_rl}} & Reinforcement learning over a knowledge graph for text communication \\\cline{2-3} 
                                                & Lotfi \etal {\csname @fileswfalse\endcsname\cite{lotfi_collab_rl}} & Collaborative deep reinforcement learning with heterogenous agents \\\cline{2-3} 
                                                & Yun \etal {\csname @fileswfalse\endcsname\cite{yun_rl_semcomm}} & Reinforcement learning for air-to-ground semantic communication \\\hline 
    \end{tabular}
    \label{tab_ml_summary}
\end{table*}
}

Many of the works previously discussed combined KG methods with DL methods, which requires an embedding of the KG into some vector space. This embedding essentially aims to preserve the knowledge represented by the graph in the embedding space. Similar to reasoning, this task becomes difficult at large scales. \cite{zheng_KG_embed} studies the problem of training KG's at scale, proposing a technique termed DGL-KE to efficiently perform KG embeddings. DGL-KE provides optimized embeddings for three types of hardware configurations: (1) many-core CPU machines, (2) multi-GPU machines, and (3) a cluster of CPU/GPU machines. For each hardware type, the DGL-KE takes advantage of parallel processing to fully utilize the computing hardware. The allocation of memory resources throughout the process is specifically designed for each hardware type, and mini-batch training is utilized to perform the embedding. Other optimization techniques employed by DGL-KE are graph partitioning, negative sampling, data access to relation embeddings, and applying gradients to global embeddings. Experiments for hardware types (1) and (2) demonstrate improved efficiency compared to other methods.

Another interesting problem is that of KG fusion. For example, say that when a new agent joins the network, we wish for it's knowledge to be merged with the overall knowledge of the network. If both knowledge bases are represented by KG's, how should we fuse them together? One way is by \textit{instance matching}, which establishes a semantic link between instances in KG's. \cite{zhu_KF_fuse} proposes a method called Follow-the-Regular-Leader Instance Matching (FTRLIM), which is able to match instances between large-scale KG's with approximately linear time complexity. The FTRLIM framework is based on an blocking algorithm called MultiObJ, which divides instances into blocks and is also developed in \cite{zhu_KF_fuse}. Through various experiments, FTRLIM is shown to perform effective and scalable KG fusion. 

For more information on the field of KG's, we direct the interested reader to \cite{ji_kg_survey}, which provides a recent and comprehensive survey covering (1) KG representation learning, (2) knowledge acquisition and completion, (3) temporal KG's, and (4) knowledge-aware applications (such as semantic communication). 

\subsection{Summary}
Following the large amount of research that has been dedicated toward the semantic web, there have been many works which seeks to utilize KG's for semantic communication. This idea stems from the fact that some kind of knowledge representation is required for semantic communication, and a common way of representing knowledge is with KG's. Furthermore, some have proposed similarity measures for concepts within a KG, which can be used to quantify semantic similarity. This form of knowledge representation has been proposed for use in the so-called semantic sensor web, which extends the semantic web to physical sensor networks. Various works have attempted to leverage KG's for different semantic communication scenarios, including recommendation systems and general goal-oriented communication. Methods such as hybrid ML-KG systems, game theoretic techniques, and others have been applied to achieve this communication. Finally, some have studied the specific problem of working with the KG itself, which will be important when implementing these systems at scale. The key idea behind KG-based semantic communication is that knowledge, and therefore meaning, can be captured by a KG and utilized by the semantic system.

Some of the works mentioned above have provided quantitative results demonstrating improved performance with regards to communication efficiency. As mentioned, results in \cite{chun_sem_iot} indicate a reduction of around 40\% in overall network traffic. Results from \cite{jeong_speech} show decreased error rates using semantic techniques, which in turn has implications for reduced communication traffic. Experiments in \cite{he_KG_agents} show that goal-oriented semantic communication can reduce the amount of entities and attributes communicated by 47\% and 17\% respectively. In a more specific example, results from \cite{aumayr_KG_recommender} demonstrate that semantic communication can reduce the amount of total information conveyed in a recommendation system by up to 91\%. These results, stemming from diverse examples and use-cases, show that KG-based semantic communication can address the issue of increasing data demands with more efficient communication.

\subsection{Challenges and Opportunities}

KG's are a popular way of representing knowledge in a system, which can then be used to facilitate semantic communication. However, this approach does not come without its challenges. First of all, as knowledge in the system grows, the KG can become massive. As was discussed in the previous subsection, working with KG's becomes difficult as they grow larger. Scalability is a challenge that must be addressed for efficient semantic communication. This difficulty is amplified in systems with stringent communication requirements, such as those requiring real-time operation. Another challenge is that a KG must be predefined with some prior knowledge of the system, reducing the ease of deployment. Building these graphs can be time-consuming and therefore costly.

These challenges present opportunities as well. Further development of scalable methods, such as those in \cite{wei_KG_reasoning, zheng_KG_embed, zhu_KF_fuse}, will be required for the practical use of large (and thus more expressive) KG's. Furthermore, techniques such as transfer learning present a promising approach toward reusing existing KG's for new applications, such that a KG does not need to be built from scratch for each deployment. Another exciting field is of ML and DL on graphs. These techniques can be employed to learn optimal embeddings and relations from existing data, and bring with them all of the benefits that have been achieved with DL in other domains.


\section{Machine Learning-Based Semantic Communication}
\label{subsec_ml}

The field of ML has seen an explosion of activity in recent years. At its core, ML seeks to learn from data in order to better perform some task. The availability of massive amounts of data and advanced algorithms have enabled the practical implementation of ML in many domains, including NLP \cite{torfi_nlp_survey}, computer vision \cite{computer_vision_survey}, and others. Over the past few years, some have sought to utilize the power and flexibility of ML to develop semantic communication systems. The general idea behind these approaches is to ``learn'' the semantics of the problem. Just as in image processing, where the key features for classifying an image may be hidden to us, the semantic aspects of communication may be unknown. Through the use of ML methods, these latent semantic features can be learned through training and added to the system's knowledge base automatically. Thus, by utilizing learning methods, these systems address one of the key challenges inherent to KG-based semantic communication, namely the requirement of a predefined knowledge base. In this section, we will take a closer look at some of these works to illustrate the ML-based approach to semantic communication. The works discussed in this section are summarized in Table \ref{tab_ml_summary}.

\subsection{Deep Learning Methods}

DL is a subset of ML which utilizes \textit{deep neural networks} (DNN's) to perform prediction and decision-making tasks \cite{dong_deeplearning_survey}. These networks are trained through an iterative update of the network parameters, which is typically accomplished through some gradient-based method. Modern DNN's come in many different forms, such as the classic multi-layer perceptron \cite{rosenblatt_1962}, convolutional neural network (CNN) \cite{cnn_survey}, recurrent neural network (RNN) \cite{rnn_survey}, etc. As with any DL problem, a key consideration is how to choose the \textit{loss function}, which will determine how the parameters are tuned. For more on information on DL, please see \cite{dong_deeplearning_survey}.

Regarding semantic communication, DL can be used in many different ways, often depending on the modality of communication (text, images, speech, etc.). One of the first works employing DL for semantic communication provides an illustrative example. In \cite{lu_radiotelephony}, aviation radiotelephony communication (ARC) is considered, where a statement spoken by one party is repeated by the other to ensure understanding. In this scenario, clear and reliable communication is crucial, as misunderstandings can lead to accidents and crashes. To minimize semantic errors (i.e., misunderstandings between pilots and air traffic control), a long short-term memory RNN (LSTM-RNN) is proposed. The LSTM-RNN takes as its input a pilot-air traffic control sentence pair. To train the network, statement pairs are assigned a similarity $R$, ranging from $R = 0$ to $R = 1$, with 1 indicating the strongest semantic similarity. Clearly, a strong similarity is desired; we want the pilot and air traffic control to be on the same page. The network is then trained to minimize the \textit{cross-entropy} error:
\begin{equation}
    \label{eq_CE_loss_fn}
    L = -\sum_{n=1}^N R^*\log(R) + (1-R^*)\log(1-R)
\end{equation}
where $R^*$ is the true label value and $R$ is the predicted consistency value. After training, the network can be used to classify new statement pairs; if a pair is found to be semantically inconsistent, a signal can be given to either party to correct the misunderstanding.

\begin{figure}[htbp]
    \centering
    \includegraphics[scale = 0.33]{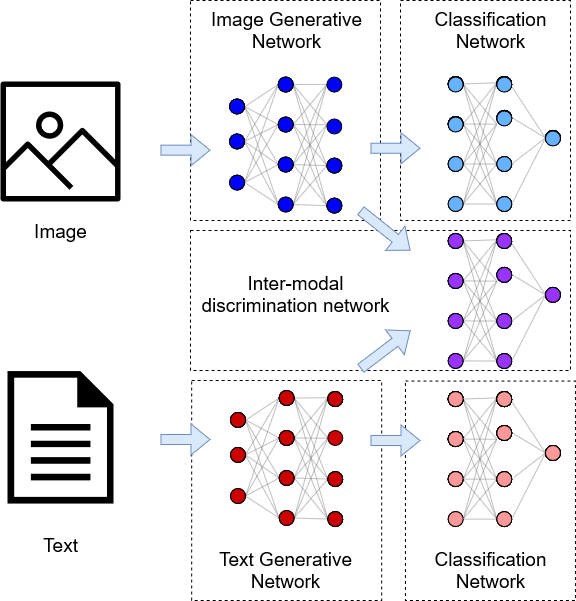}
    \caption{Basic framework of the multi-modal GAN approach proposed in \cite{hua_deep_sem_corr}}
    \label{fig_multimodal_gan}
\end{figure}

Another interesting problem involving semantic communication is that of cross-modal retrieval, in which the format of the query is different than that of the information being queried, e.g., a voice request for certain textual document in a database. In \cite{hua_deep_sem_corr}, a generative adversarial network (GAN) is proposed to address this problem by performing \textit{semantic correlation} on multi-modal data, specifically image and text data. A GAN is a network composed of two networks, namely a generator and a discriminator. The generator learns by attempting to ``fool'' the discriminator, while the discriminator is optimized to discern between outputs of the generator. \cite{hua_deep_sem_corr} proposes a framework in which the generator takes both an image and a body of text as inputs and learns their representations. The goal of the discriminator is to distinguish between the two modalities. The basic framework of this approach is illustrated in Figure \ref{fig_multimodal_gan}. After training, the generator will have learned the representations of the heterogeneous modalities in a common space, in which data with similar semantics will be close, i.e., the \textit{meaning} common to both the text and the image is captured in this space. The proposed method is shown to outperform both traditional and deep methods with respect to the mean average precision metric.

Another work utilizing a GAN is \cite{huang_dl_image_coding}, in which DL is implemented with the goal of semantic coding of images. The goal of semantic coding is to minimize the bit rate of transmission while preserving the semantic information of the image. Again, the semantics of interest here are unknown, and the goal is to learn them throughout the training process. Here, the generator network is used to learn and restore semantic information which is used as a ``base layer'' of the image. The generator loss function is formed as a rate-perception-distortion trade-off, including a combination of the VGG loss \cite{ledig_vgg_loss} and LPIPS \cite{halevy_lpips_loss}. Then, Better Portable Graphics (BPG) residual coding is used to refine the image. The overall network explores different strategies to optimize the rate-perception-distortion tradeoff, and is shown to exhibit similar performance to baselines models while utilizing a 2-4$\times$ reduced bit rate.

\begin{figure*}[htbp]
    \centering
    \includegraphics[scale = 0.27]{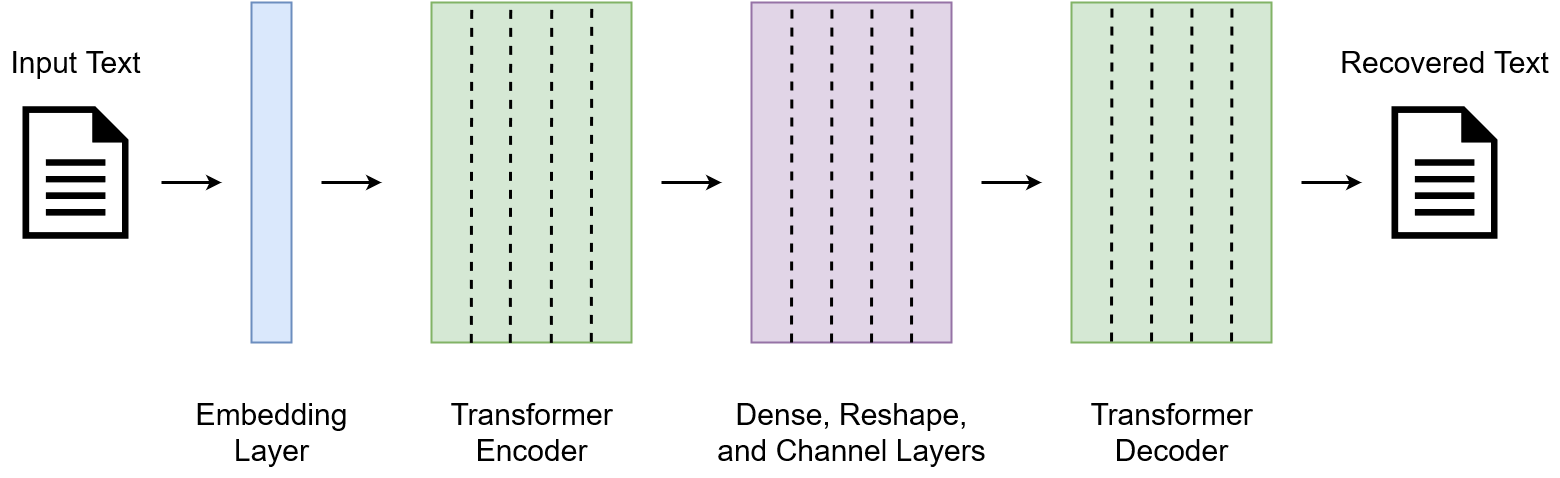}
    \caption{Architecture of the end-to-end semantic communication system DeepSC proposed in \cite{xie_deepsc}. Each box represents a different section of the network, which are all trained in a joint manner.}
    \label{fig_deepsc_diagram}
\end{figure*}

Similar to \cite{zhang_ernie}, \cite{qiao_seed} also looks to enhance standard encoder-decoder NLP techniques (this time without the use of a KG), specifically for the problem of recognizing some text within an image, or \textit{scene text recognition}. For example, in the case of a self-driving vehicle, it would be beneficial for the vehicle to be able to recognize the text located on street signs within the field of vision, as a human driver would. In their model, termed SEED, semantic information is used in both the encoder module for supervision and the decoder module for initialization. The semantic information is predicted from the image features which are first extracted with a CNN and RNN. To predict this information, a simple fully-connected DNN is trained with a dual cross-entropy and cosine embedding loss function. Then, the image features and the semantic information are both fed to the decoder to perform text recognition. Using ASTER \cite{baoguang_aster} as an exemplar to demonstrate the framework, it is shown that the semantic enhancement improves performance of the model.

A CNN along with federated learning (FL) is proposed to facilitate audio-based semantic communication in \cite{tong_audio_semcomm}. FL is a branch of DL in which distributed, local models are trained individually and then combined to form a global model. In \cite{tong_audio_semcomm}, FL is implemented in a system consisting of a single server and many devices. The devices train local models based on data, to reduce communication overhead to the server. Each local model consists of an autoencoder with a convolutional layer for extracting the semantic information of the audio. Normalized root mean square error is used as a loss function to evaluate the quality of semantic reconstruction, and experiments show that the proposed architecture can achieve around 100$\times$ improved performance (with respect to the mean square error) with around 1/3 the transmitted data of traditional methods.

Yet another type of DNN is the \textit{transformer}, which has seen wide success in the field of NLP \cite{torfi_nlp_survey}. Transformers rely on the idea of \textit{attention}, which provides different weights to different features of the input data, similar to how our brains pay more attention to certain perceptual inputs. In \cite{zhou_universal_tx}, an adaptive universal transformer is implemented for text-based semantic communication. The optimization goal is ``to minimize the semantic errors while facing different communication situations.'' The universal transformer is able to accomplish this by adding a circulation mechanism which can dynamically allocate greater computation time to semantically complex statements. Intuitively, this is similar to how a human might read some text. Passages with a simple meaning are easier to understand and thus can be read faster, while passages with complex meaning demand more thought (equivalently, computation time). Cross-entropy is used as the loss function for training the network, and simulations performed on the standard proceedings of the European Parliament \cite{koehn-2005-europarl} show an improved performance over traditional methods.

The transformer has also been proposed for semantic communication in \cite{sana_semcomm_6G}. In this work, an end-to-end architecture is proposed which performs communication with \textit{semantic symbols}, which are used to represent semantics. The semantic communication model is defined, including the encoder, decoder, channel and noise. The transformer model is then designed for this model, including both source and channel coding in a joint architecture. This architecture is trained using the cross-entropy loss, and experiments in a NLP setting demonstrate that the proposed system can achieve a similar performance to traditional techniques with a 21\% decrease in the number of symbols.

\subsubsection{DeepSC and its Variants}

One particular DL model for semantic communication that has been the subject of various studies was first proposed in \cite{xie_deepsc}, where it was given the name DeepSC. DeepSC is also based on the transformer DL model, and the model proposed in \cite{zhou_universal_tx} takes its inspiration from DeepSC. Similar to other works, \cite{xie_deepsc} defines the model for the semantic communication system, which consists of a transmitter performing both semantic encoding and channel coding, and a receiver performing the inverse operations. The architecture of the proposed network is shown in Figure \ref{fig_deepsc_diagram}. Both transmitter and receiver possess some background knowledge. The goal is stated as simultaneously minimizing semantic errors (measured with the cross-entropy loss function) and transmitted symbols. This is accomplished through an end-to-end transformer network, which uses a self-attention mechanisms for extracting semantic information from text; here, the meaning is captured by this attention mechanism and which input text is emphasized by the model. Various metrics are used to demonstrate the superior performance of DeepSC compared with traditional communication methods.

There have been a few variants inspired by DeepSC, which all aim to facilitate semantic communication within different modalities. One example is \cite{xie_deepsc_speech}, which proposes a semantic communication system for speech signals, termed DeepSC-S. It is claimed that the end-to-end communication system ``learns and extracts the essential speech information." This is a very intuitive idea, as it is clear how non-verbal qualities of speech can impact a conversation (tone, volume, etc.). Specifically, DeepSC-S employs an attention-based CNN for speech coding and a CNN for channel coding, and the mean-squared error loss function is used for training. It is confirmed through simulations that DeepSC-S outperforms traditional systems with equivalent bit rates under AWGN, Rayleigh and Rician channels.

{\renewcommand{\arraystretch}{1.3}
\begin{table*}[htbp]
    \centering
    \caption{Summary of DL-based Semantic Communication Loss Functions}
    \begin{tabular}{|p{.18\linewidth}|p{.42\linewidth}|p{.32\linewidth}|}\hline
    \multirow{1}{\linewidth}{\centering \textbf{Loss Function}} & \multirow{1}{\linewidth}{\centering \textbf{Expression}} &  \multirow{1}{\linewidth}{\centering \textbf{Description}} \\\hline
    \multirow{2}{\linewidth}{\centering Cross-entropy {\csname @fileswfalse\endcsname\cite{lu_radiotelephony, zhou_universal_tx, sana_semcomm_6G, xie_deepsc, xie_deepsc_vqa, xie_lite_deepsc}}} & \multirow{2}{\linewidth}{\centering $ L = -\sum_{n=1}^N R^*\log(R) + (1-R^*)\log(1-R)$} &  \multirow{2}{\linewidth}{\centering Promotes semantic similarity or probability values $R$ consistent with labels $R^*$} \\
    & & \\\hline
    \multirow{4}{\linewidth}{\centering Cross-modal generative {\csname @fileswfalse\endcsname\cite{hua_deep_sem_corr}}}  & \multirow{4}{\linewidth}{\centering $L_{gen} = \lambda L_{adv} + L_{cls}$, $L_{adv} = -\sum_{i=1}^N \left[ \log(1 - D(G_I(x_i))) + \log D(G_T(y_i)) \right]$, $L_{cls} = -\sum_{i=1}^N\log \left[ C(G_I(x_i))^Tc_i + \beta C(G_T(y_i))^Tc_i \right]$} & \multirow{4}{\linewidth}{\centering Aims to promote similar image embeddings $G_I(x_i)$ and text embeddgins $G_T(y_i)$ given semantically similar inputs $x_i$ and $y_i$} \\
    & & \\
    & & \\
    & & \\\hline
    \multirow{3}{\linewidth}{\centering Rate-perception-distortion {\csname @fileswfalse\endcsname\cite{huang_dl_image_coding}}} & \multirow{3}{\linewidth}{\centering $\mathcal{L}_{E,G} = \mathcal{L}_{G} + \lambda_1 (\mathbb{E}[d(x,G(\hat{w},s))] + \mathbb{E}[\Vert G(\hat{w},s) - x\Vert_1 ] ) + \lambda_2(R(\hat{w}) + R(s) + R(r'))$} &  \multirow{3}{\linewidth}{\centering Simultaneously minimizes the rate, perception and distortion with hyperparameters $\lambda_1$ and $\lambda_2$} \\
    & & \\
    & & \\\hline
    \multirow{2}{\linewidth}{\centering Cross-entropy/cosine embedding {\csname @fileswfalse\endcsname\cite{qiao_seed}}} & \multirow{2}{\linewidth}{\centering $L = L_{rec} + \lambda [1-\cos(S,em)]$} &  \multirow{2}{\linewidth}{\centering Combination of traditional cross-entropy $L_{rec}$ and cosine embedding loss} \\
    & & \\\hline
    \multirow{2}{\linewidth}{\centering Normalized root mean squared error {\csname @fileswfalse\endcsname\cite{tong_audio_semcomm}}} & \multirow{2}{\linewidth}{\centering $\mathcal{L}_{NRMSE} = \frac{\sum_{t=1}^T (a_t - \hat{a}_t)^2}{\sum_{t=1}^Ta_t^2}$} &  \multirow{2}{\linewidth}{\centering Seeks to minimize the difference between recovered audio data $\hat{a}_t$ and actual data $a_t$} \\
    & & \\\hline
    \multirow{2}{\linewidth}{\centering Mean squared error {\csname @fileswfalse\endcsname\cite{xie_deepsc_speech}}} & \multirow{2}{\linewidth}{\centering $ \mathcal{L}_{MSE} = \frac{1}{W}\sum_{w=1}^W (s_w - \hat{s}_w)^2$} &  \multirow{2}{\linewidth}{\centering Seeks to minimize the difference between recovered speech data $\hat{s}_w$ and actual data $s_w$} \\
    & & \\\hline
    \end{tabular}
    \label{tab_ml_losses}
\end{table*}
}

Another variation of DeepSC is proposed in \cite{xie_deepsc_vqa}, which considers a multimodal communication system and is termed MU-DeepSC. The task-oriented system is implemented for visual question answering (VQA), where the transmitter sends an image and a question about that image (text) and the receiver aims to correctly answer the question (e.g., is there a red ball present?). The transmitter consists of two networks; one employs ResNet-101 \cite{resnet101} and a CNN for semantic image encoding, and the other utilizes Bi-LSTM \cite{bi_lstm} and a dense DNN for semantic text encoding. These encodings are transmitted over the channel, where the receiver then implements a memory, attention, and composition (MAC) network to generate the answer. The end-to-end network is trained with the cross-entropy loss, and simulation results using the CLEVR dataset \cite{johnson_clevr_dataset} demonstrate accuracy gains of up to 80\% compared to traditional methods with a 70\% reduction in transmitted symbols. 

As we've seen in the discussion of the semantic sensor web, a prevalent idea is the use of semantic communication in IoT applications. In \cite{xie_lite_deepsc}, a ``lite'' version of DeepSC, named L-DeepSC, is proposed for use with IoT networks. In L-DeepSC, the semantic communication model is trained and updated in the cloud and distributed to IoT devices. These devices then implement the model to perform semantically aware data collection and text transmission with low complexity. The system first uses a least-squares estimator to obtain channel state information (CSI), and then a deep de-noise network to refine the CSI estimates. The trained model is then compressed through sparsification and quantization and broadcast to the IoT devices. The devices then use this model to perform semantic communication with text data, uploading new data to the cloud. It is shown that the proposed system performs competitively with traditional methods, especially in the low SNR domain. Moreover, L-DeepSC reduces the model parameters of the original DeepSC network by around 60\%, which translates to reduced communication upon model distribution to IoT devices.

As ML-based semantic communication is based on \textit{learning} the semantics of the problem, explicit semantic metrics are not defined. However, the chosen \textit{loss function} affects how the semantics are learned. Table \ref{tab_ml_losses} provides the loss functions used in each of the discussed DL approaches.

\subsection{Reinforcement Learning Methods}

\begin{figure}[htbp]
    \centering
    \includegraphics[scale = 0.45]{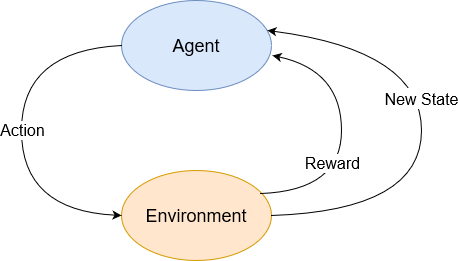}
    \caption{Reinforcement learning framework}
    \label{fig_RL}
\end{figure}

Another popular approach to ML is reinforcement learning (RL) \cite{rl_survey}. In reinforcement learning, the model is viewed as an agent in a state space. From its current state, the agent can take some action, for which it is provided a reward. The goal of the agent is to discover the action-taking policy which will maximize long-term rewards from any given state. Rather than use existing data to determine this policy, in RL the model is trained by letting the agent ``explore'' different policies; the model takes some sequence of actions, and then tunes the parameters based on the rewards received. A simple illustration of this learning process is given in Figure \ref{fig_RL}. Due to the agent/reward set up of RL, it is a method which is well-suited for learning to play different games, such as chess and Go. Indeed, learning models including both RL and DL have been used to create artificial players which outperform human world champions in both games \cite{ai100}.

Recently, some have turned to using RL as a means to achieve semantic communication. In \cite{lu_rl_semcomm}, a RL solution is proposed to carry out general semantic communication. It is argued that the objective functions of many of the ML-based approaches to semantic communication demonstrate a ``semantic blindness,'' and are still biased toward bit-level accuracy. The proposed joint source-channel coding solution, termed SemanticRL-JSCC, is formulated using a Markov decision process framework. The reward function is based on any \textit{general} semantic similarity function. This method differs from those DL methods discussed, as the semantics of the problem are defined by the chosen similarity function, and thus SemanticRL focuses on \textit{how} to best communicate given some semantics, rather than learning \textit{what} those semantics are. A distinct feature of SemanticRL as opposed to other ML approaches is that this similarity function is not necessarily differentiable. In the training of the model, a \textit{self-critic} approach \cite{rennie_selfcritic_rl} is taken, resulting in a quicker and simpler solution. Experiments carried out over the European parliament dataset demonstrate superior performance with regard to common metrics, as well as a stable learning trajectory.

Another work combines a KG representation of semantic information with RL for semantic communication using text data \cite{wang_kg_rl}. First, a KG is extracted from a body of text, and this KG is treated as the semantic information of that text. Based on this KG representation, two metrics, namely semantic \textit{accuracy} and \textit{completeness} are derived. Combining these two metrics gives the overall metric of semantic similarity. Semantic communication is then formed as an optimization problem which seeks to maximize the semantic similarity of the text at the transmitter and receiver through resource allocation and information transmission, under a delay constraint. Using an attention-based RL framework to solve this optimization problem, it is shown that the proposed semantic communication solution outperforms a traditional RL scheme as well as typical wireless communication techniques.

{\renewcommand{\arraystretch}{1.5}
\begin{table*}[t]
    \centering
    \caption{Summary of Works in Significance-Based Semantic Communication}
    \begin{tabular}{|c|c|c|}\hline
    \multirow{4}{*}{Age of Information}     & Uysal \etal {\csname @fileswfalse\endcsname\cite{uysal_semcomm_2021}} & Age of information as a significance-based semantic metric \\\cline{2-3} 
                                            & Uysal \etal {\csname @fileswfalse\endcsname\cite{uysal_aoi_practice}} & Practical evaluation of the age of information metric \\\cline{2-3} 
                                            & Beytur \etal {\csname @fileswfalse\endcsname\cite{beytur_tcp_vs_udp_aoi}} &  Age of information performance of UDP and TCP protocols \\\cline{2-3} 
                                            & Ayan \etal {\csname @fileswfalse\endcsname\cite{ayan_aoi_vs_voi}} & Comparison of AoI and VoI impact on performance of cellular control system \\\hline 
    \multirow{3}{*}{Value of Information}   & Uysal \etal {\csname @fileswfalse\endcsname\cite{uysal_semcomm_2021}} & Value of information as a significance-based semantic metric \\\cline{2-3} 
                                            & Molin \etal {\csname @fileswfalse\endcsname\cite{molin_voi}} & Value-based information management for state estimation \\\cline{2-3} 
                                            & Ayan \etal {\csname @fileswfalse\endcsname\cite{ayan_aoi_vs_voi}} & Comparison of AoI and VoI impact on performance of cellular control system \\\hline 
    \multirow{4}{*}{Semantic Sampling}      & Uysal \etal {\csname @fileswfalse\endcsname\cite{uysal_semcomm_2021}} & Semantic sampling as part of a semantic communication architecture \\\cline{2-3} 
                                            & Kountouris \& Pappas {\csname @fileswfalse\endcsname\cite{kountouris_2021}} & Semantic sampling as part of a semantic communication architecture \\\cline{2-3} 
                                            & Bacinoglu \etal {\csname @fileswfalse\endcsname\cite{bacinoglu_semsamp}} & Semantic sampling for tracking of a stochastic process \\\cline{2-3} 
                                            & Dommel \etal {\csname @fileswfalse\endcsname\cite{dommel_semsamp}} & Semantic sampling for goal-oriented communication over a shared medium \\\hline 
    \end{tabular}
    \label{tab_sig_summary}
\end{table*}
}

{\renewcommand{\arraystretch}{1.3}
\begin{table*}[!b]
    \centering
    \caption{Summary of Significance-Based Semantic Measures \cite{uysal_semcomm_2021}}
    \begin{tabular}{|p{.22\linewidth}|p{.21\linewidth}|p{.50\linewidth}|}\hline
    \multirow{1}{\linewidth}{\centering \textbf{Measure}} &  \multirow{1}{\linewidth}{\centering \textbf{Expression}} &  \multirow{1}{\linewidth}{\centering \textbf{Description}} \\\hline
    \multirow{1}{\linewidth}{\centering Age of Information} & \multirow{1}{\linewidth}{\centering $\Delta(t) = t - u(t)$} &  \multirow{1}{\linewidth}{\centering Characterizes the \textit{freshness} of information generated at time $u(t)$} \\\hline
    \multirow{1}{\linewidth}{\centering Value of Information}  & \multirow{1}{\linewidth}{\centering $v(x) = b(x) - c(x)$} & \multirow{1}{\linewidth}{\centering Difference between the benefit of a sample and cost of its transmission} \\\hline
    \multirow{1}{\linewidth}{\centering Relevance of Information} & \multirow{1}{\linewidth}{\centering $r(x) = f(x_n,x_{n-1})$} &  \multirow{1}{\linewidth}{\centering Quantifies the amount of change in a process since the previous sample} \\\hline
    \end{tabular}
    \label{tab_sig_measures}
\end{table*}
}

Collaborative RL is a form of RL in which multiple agents are present in the system, and they collaborate to determine the optimum policy. In \cite{lotfi_collab_rl}, collaborative deep RL (CDRL) is used to train a group of heterogeneous agents over a wireless cellular network. First, the algorithm selects the best subset of semantically relevant DRL agents for collaboration.  This semantic relevance between two agents is based on their policies; if a target agent returns a large average reward under a source agent's policy, the target is said to be similar to the source. Here, the semantics are captured by the policies of each of the agents; similar meaning is implied by a similar policy. Once the similar subset of agents is obtained, the training loss and wireless bandwidth are jointly minimized to obtain the optimal policies for each agent. Simulations of the proposed technique show improved training performance compared to other CDRL methods and classic DRL. It is also shown that the proposed approach is able to use resources more efficiently, demonstrating better performance with fewer resource blocks than other approaches.

Looking to implement RL-based semantic communication for a specific application, \cite{yun_rl_semcomm} proposes a DRL framework for air-to-ground URLLC communcation using unmanned aerial vehicles (UAV's). Similar to \cite{lotfi_collab_rl}, this work proposes the use of a multi-agent DRL framework, coined graph attention exchange network (GAXNet), for semantic communication. Self-attention is used to determine the attention a UAV gives to other UAV's in the network, and based on this attention, training is performed in a centralized manner. Once the optimal policies have been obtained, the central unit distributes the model to the UAV's and actions are carried out in a decentralized manner. It is shown that the proposed GAXNet achieves more efficient training than the state-of-the-art centralized training and decentralized execution algorithm QMIX \cite{rashid_qmix}, and is better able to avoid collisions between UAV's.

\subsection{Summary}
ML-based semantic communication is an approach that has seen a spike of interest with the recent boom in AI technology. In ML-based semantic communication, the semantics of the problem are not predefined as in classical and KG-based semantic communication, but rather they are learning through data-driven training. This learning is performed either through DL or RL. Many different DL models have been proposed to facilitate semantic communication, include the transformer, GAN, CNN, and others. One notable example is DeepSC, which was originally proposed as a text-based semantic communication system using a transformer-based network. Variants of DeepSC have been proposed in recent years which target different modes of communication. Though not as prominent, some RL methods have been proposed to learn semantics as well. One benefit of this approach is the loss function need not be differentiable. In ML-based semantic communication, meaning is characterized by the parameters of a model which are learned in a data-driven manner.

Many of the discussed works provide quantitative results demonstrating efficient semantic communication. The results of \cite{huang_dl_image_coding} show a 2-4$\times$ reduction in bit rate, while experiments in \cite{tong_audio_semcomm} indicate large performance gains over traditional systems for around 1/3 the original bit rate. \cite{sana_semcomm_6G} demonstrates a 20\% reduction in transmitted symbols to achieve a similar performance as baseline systems. DeepSC and it's variants also indicate potential for traffic reduction; DeepSC-S \cite{xie_deepsc_speech} can achieve improved speech performance for similar bit rates, and MU-DeepSC \cite{xie_deepsc_vqa} achieves superior accuracy with a 70\% reduction in transmitted symbols. Finally, L-DeepSC \cite{xie_lite_deepsc} reduces traffic in another way, by drastically reduce the parameters of a ML model which is to be broadcast to IoT devices. All in all, as in KG-based semantic communication, existing works in ML-based semantic communication indicate the potential for this approach to address the issues raised at the outset.

\subsection{Challenges and Opportunities}

One of the challenges with ML-based approaches to semantic communication was pointed out in \cite{lu_rl_semcomm}, which is the difficulty in working with semantic metrics. As most DL techniques use gradient-based methods for optimization, we must necessarily work with differentiable metrics, while many semantic similarity metrics are non-differentiable. Another critical challenge is the black-box nature of many DL models. A challenge within the field of DL as a whole, this quality of deep networks obscures our ability to analyze and evaluate \textit{why} a model does or does not perform well. Finally, the improved performance that DL has enjoyed due to big data is approaching a limit, where futher gains can only be reached through massive training, which imposes a huge computational burden \cite{shlezinger_modelbaseddl}.

Opportunities can be found in the solutions to these challenges. First, there is certainly no ubiquitous metric of semantic similarity, and it is likely that many metrics will be application-specific. Development of such metrics is of critical importance to the advancement of semantic communication. As each approach to semantic communication we've seen thus far (classical, KG-based, ML-based) employs its own unique metrics, it is likely that a convergence of these ideas and combination of metrics could yield novel implementations and improved results. Furthermore, development of methods similar to \cite{lu_rl_semcomm} that relax the differentiability requirement of the semantic metric is a promising avenue for further study.

Regarding the lack of interpretability and ever-growing appetite for data associated with deep networks, one promising solution is the development of \textit{model-based deep learning} techniques \cite{shlezinger_modelbaseddl}. Model-based methods are those that derive some inference rule based on prior knowledge of the problem, while data-driven methods (including DL) rely solely on data to form the inference rule. Model-based methods are typically more interpretable and efficient than data-driven methods, while data-driven methods are more expressive and robust in new situations. Techniques utilizing the strengths of both methods to create a single model are referred to as model-based DL \cite{shlezinger_modelbaseddl}, and are another promising method of addressing some of the challenges facing ML today.

\section{Significance-Based Semantic Communication}
\label{subsec_right}

\begin{figure*}
    \centering
    \includegraphics[scale = 0.3]{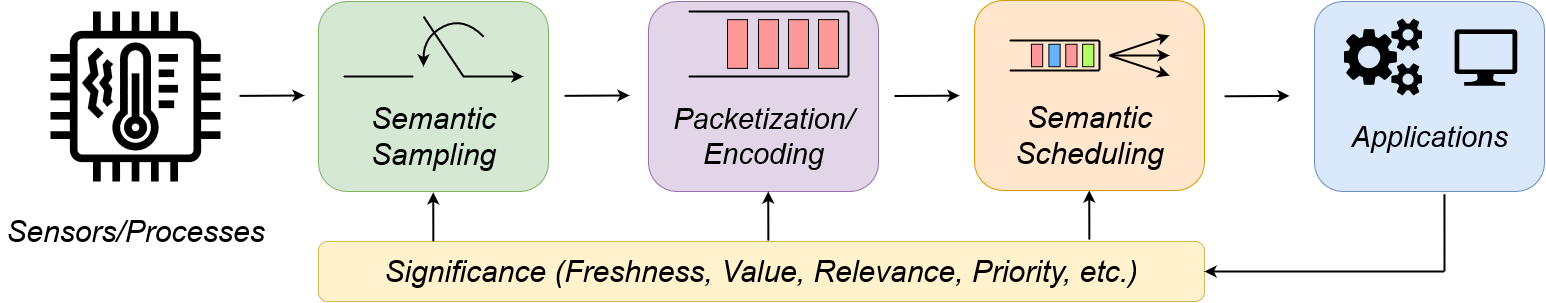}
    \caption{Overview of the significance-based semantic architecture proposed in \cite{uysal_semcomm_2021}}
    \label{fig_sig_based_flowchart}
\end{figure*}

One last view on semantic communications has been proposed only recently, and involves defining the semantics of information as the \textit{significance} of this information \cite{uysal_semcomm_2021}. Recall ``The effectiveness problem" defined in Section \ref{sec_intro}: ``How effectively does the received meaning affect conduct in the desired way?" Defining the semantics of information as significance of information essentially addresses this problem, as significance is inherently determined by what one is trying to achieve with communication, or the \textit{goal}. As a simple example, if the goal of communication is to control a robot performing remote surgery (a necessarily real-time application), information that was just obtained will be much more significant than information obtained 10 seconds ago. Based on this general idea, \cite{uysal_semcomm_2021} calls for ``a redesign of the entire process of information generation, transmission and usage in unison.'' In this section, we survey the few recent works that support this idea of significance-based semantic communication. The works discussed in this section are summarized in Table \ref{tab_sig_summary}.

In \cite{uysal_semcomm_2021}, some examples of measures which relate to the significance of information are given. The first of these is information \textit{freshness}, which is determined by the time taken since the information was generated to when it was received. Age of Information (AoI) is a measure which captures this idea of freshness, and has been well-studied over the past decade or so. We will discuss AoI further in the following subsection as a prime example of a significance-based semantic measure; \cite{uysal_semcomm_2021} presents results indicating energy savings of different age-aware protocols ranging from 10-64\%. Another example of a semantic measure is \textit{relevance}. Consider a process that is being sampled; consecutive samples that capture little change in the process are typically of less interest than those for which sudden changes occur. We could say that the latter samples are more relevant than the former. As an extension of relevance of information, a more powerful example is given as the \textit{value} of information, which is defined as the difference between the benefit of a sample and the cost of its transmission. It is argued in \cite{uysal_semcomm_2021} that developing metrics that capture these ideas is critical for achieving semantic communication. Those mentioned here are summarized in Table \ref{tab_sig_measures}.

\cite{uysal_semcomm_2021} also presents a vision for an end-to-end semantic communication architecture. This architecture takes into considerations the elements of freshness, relevance, and value to optimize the entire system. A simplified flowchart of the proposed architecture is provided in Figure \ref{fig_sig_based_flowchart}. Specifically, semantic sampling is implemented to relax the assumption that data arrives in an uncontrolled manner, i.e., only significant information is generated in the first place. Semantic channel encoding, multiuser scheduling, channel access and flow control are proposed to increase the efficiency and effectiveness of each of these processes. It is stated that the realization of this architecture will involve an entirely new paradigm shift that is incompatible with previous designs of communication systems. A few specific examples are envisioned, which include semantic communication for networked control systems, smart cities, and mMTC/IoT systems.

\cite{kountouris_2021} is a seemingly independent work from that of \cite{uysal_semcomm_2021} which emerged around the same time and shares the idea of significance-based semantic communication. \cite{kountouris_2021} defines semantics of information as ``the significance and usefulness of messages.'' A similar argument to that of \cite{uysal_semcomm_2021} is given, stating that simple generation and communication of data often leads to reception of stale or irrelevant information at the receiver and wasted resources. This again brings about the need for a \textit{goal-oriented} communication system, one that addresses ``The effectiveness problem.''

A new concept presented in \cite{kountouris_2021} is the idea of defining semantics at different \textit{scales}. At the \textit{microscopic} scale, specific pieces of information from the source may be of different significance, e.g., that a safety risk is present or not. The \textit{mesoscopic} scale is the intermediate level, which takes into consideration link-level semantics. This includes both innate (objective) measures such as freshness (AoI) and precision, and contextual (subjective) measures such as timeliness and completeness. Finally, the \textit{macroscopic} scale takes system-level semantics into consideration, specifically looking at end-to-end distortions and delays that affect the end goal.

An end-to-end semantic architecture is proposed, which is strikingly similar to that envisioned in \cite{uysal_semcomm_2021}, including semantic sampling and semantics-aware signal processing blocks. A specific example of an end-to-end communication system is given, involving a remote actuation application. The source monitors the actual state of a robotic arm, while the receiver aims to construct and maintain a digital twin of this robotic arm. Communicating over a wireless erasure channel, it is shown that an end-to-end semantics approach performs much better than other, more semantically-unaware approaches in terms of real-time reconstruction error and cost of actuation error.

\subsection{Age of Information and Value of Information}

Two proposed measures that capture the significance of information are AoI and VoI. In this subsection, we will discuss some of the work that has been done with regards to these two measures to illustrate their use as semantic measures. 

\begin{figure}[htbp]
    \centering
    \includegraphics[scale = 0.64]{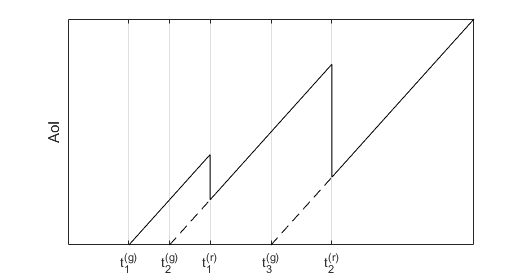}
    \caption{Example of AoI, where $t_n^{(g)}$ and $t_n^{(r)}$ is the generation and reception time of the $n$th packet, respectfully}
    \label{fig_AoI}
\end{figure}

\cite{uysal_aoi_practice} provides a compilation of some recent works examing AoI in practical scenarios and looks at issues such as synchronization, transport layer protocols, congestion, and the use of ML. First, the status \textit{age} of an information flow (characterized as a flow of data packets) is defined as the difference between the current time and the generation time of the most recently received data packet, which is illustrated in Figure \ref{fig_AoI}. Other derivative metrics are defined as well, such as \textit{average} AoI and \textit{peak} AoI. Depending on the application at hand, one may desire to minimize either the average or peak AoI to best maximize information freshness.

Next, \cite{uysal_aoi_practice} discusses the measurement of AoI in practical systems. It is noted that first and foremost, accurate timing information is required for age measurement. Second, synchronization is required at the transmitter and receiver. Once these have been established, AoI can be computed at the transmitter, receiver, or centrally and used for network optimization. It is also shown how timing imperfections, such as clock bias, can affect AoI measurement. 

\cite{uysal_aoi_practice} then looks at a specific work which examines the AoI performance of some modern transport layer protocols. In \cite{beytur_tcp_vs_udp_aoi}, the AoI performance of User Datagram Protocol (UDP) and Transmission Control Protocol (TCP) are examined for different testbed setups. It is shown that UDP is able to maintain a lower average AoI at higher data rates than TCP for a multi-hop network testbed. Both protocols tend to perform relatively well up to a certain rate, above which the system becomes ``panicked'' and AoI performance becomes poor. In contrast, evaluated with respect to an IoT tesetbed, the opposite holds true, and TCP is shown to achieve a slightly better AoI performance. Overall, \cite{uysal_aoi_practice} provides a useful summary of how AoI can be practically integrated to evaluate communication systems.

As an example application of VoI, \cite{molin_voi} proposes a value-based method of information management of a networked system for state estimation. Essentially, this system will allocate a time slot to the estimator with the highest-priority information, where priority is determined by the VoI. The VoI of each estimator is computed as a function of the expected overall weighted squared error, given the current data at that local estimator. Therefore, by choosing the data which minimizes this error, the system is essentially choosing the information with the highest value to the task at hand. To illustrate the performance of the proposed system, an automated driving scenario with multiple vehicles is simulated. It is shown that the VoI-based scheme is able to avoid collisions with very high probability, while a simple time-triggered scheduling approach resulted in a collision in 19.7\% of the experiments.

Another work has directly compared the impact of AoI and VoI on the performance of a cellular networked control system \cite{ayan_aoi_vs_voi}. Here, VoI is defined as quantifying the amount of reduction in uncertainty of a stochastic process at the recipient. It is interesting to note the similarity between this definition and that of Shannon's definition of entropy \cite{shannon}, which also quantifies uncertainty reduction. Here, VoI is concerned with the content of a new update, while AoI focuses only on the timeliness of this update. The AoI is defined in the usual fashion, while a VoI metric is proposed for both uplink and downlink transmissions, and in both cases is a function of the expected squared-error. Simulations demonstrate that a system implementing a VoI-based scheduler is able to achieve a lower absolute error as opposed to a system with an AoI-based schedule for the cellular networked control system.

\subsection{Semantic Sampling}

Another important aspect of a significance-based semantic communication system, as proposed by both \cite{uysal_semcomm_2021} and \cite{kountouris_2021} is the idea of \textit{semantic sampling}. Basically, the aim of semantic sampling is to generate information at the source in a ``smart'' way, such that only necessary information is generated and transmitted over the system. As a general example of this, \cite{bacinoglu_semsamp} considers the problem of tracking an unstable stochastic process by using causal information of another stochastic process. Essentially, by using some information related to the process of interest, we can determine when to take ``significant'' samples that allow for accurate tracking, thereby implementing semantic sampling. This work can be seen as contributing to a theory of semantic sampling as discussed in \cite{uysal_semcomm_2021}. In \cite{bacinoglu_semsamp}, necessary conditions are provided for tracking integer-valued sources using causal information. These results are expressed in terms of the R\'enyi entropy and information density; essentially, the information density between the two processes must be greater than a threshold that is set by the R\'enyi entropy of the process being tracked. Furthermore, \cite{bacinoglu_semsamp} also provides sufficient conditions for tracking integer-valued sources using causal information. The first of these is based on MAP estimator of the source information based on the causal information, and the second is based on a different estimator which considers a notion of distance. With regards to semantic sampling, the results of \cite{bacinoglu_semsamp} imply that one could perhaps only sample and transmit the causal information over the channel instead of the source information itself. If the causal information results in fewer transmitted symbols, this would theoretically increase the efficiency of our system while preserving reconstruction fidelity at the receiver.

Looking at a more practical implementation of semantic sampling, \cite{dommel_semsamp} addresses the problem of semantics-aware active sampling and transmission over a shared communication medium. The goal of the system is to use joint sampling and transmission to compute the probability of a quantity of interest at the receiver. In this work, semantics-aware communication refers to a system in which the receiver aims to recover the aggregated information of interest, rather than the individual messages. For example, perhaps the average measurement from a number of sensors is of interest; in this case, meaning is captured by the aggregated value rather than the individual data. An active sampling scheme is adopted, such that each device takes samples according to a Bernoulli distribution, where the parameter of this distribution can vary between devices. To obtain the empirical probability of the quantity of interest, each device transmits over a time slot, and the receiver averages over the time slots. The estimation technique is shown to perform well with respect to both mean squared-error and Kullback-Leibler Divergence metrics.

\subsection{Summary}
One final approach to semantic communication is significance-based communication, which addresses both the semantic and the effectiveness problems of communication. This recent approach makes use of metrics which quantify the significance of information, such as freshness, value, relevance, and others. Two prime examples of metrics corresponding to significance-based semantic communication are AoI and VoI. AoI is a popular metric which has been well-studied compared to VoI, and some work has been done comparing the two measures. Furthermore, a key idea of this approach is semantic sampling, which has received some attention as well. Overall, significance-based semantic communication looks to solve the semantic problem by first solving the effectiveness problem, and assigning meaning to information based on what impact that information will have at the receiver.

As the most recent of the discussed methods, quantitative results demonstrating the potential of significance-based semantic communication for data traffic reduction are few. In \cite{uysal_semcomm_2021}, results are presented demonstrating energy savings, resulting in decreased transmission, of 10-64\% for different age-aware protocols. While there are not many quantitative results in this area, the potential is clear. Significance-based communications center around the reduction of insignificant information, and thereby inherently work to reduce traffic in a communication system.

\subsection{Challenges and Opportunities}

One clear challenge faced by significance-based semantic communication, similar to other methods of semantic communication, is the development of applicable metrics. AoI and VoI are two examples which have been the focus of prior work, however we anticipate that other useful metrics will be proposed as this approach to semantic communication is being developed in the literature. The quest for appropriate metrics presents a rich opportunity for future research. Clearly these measures are highly application-specific, and thus the utility of one measure may vary greatly from one scenario to another.

Another challenge relates to the general progression to such a semantic communication system. As stated in \cite{uysal_semcomm_2021}, this approach entails a radical departure from the ways in which current communication systems operate. However, to be a viable path forward for modern communication systems, a certain level of backward-compatibility must be present, such that the vast existing wireless infrastructure need not be replaced from scratch. Methods which involve some degree of compatibility with legacy systems will be important for the progression to semantic communication. 

Many opportunities lie in the development of significance-based semantic communication systems. With the advancement of IoT and cyber-physical systems, wireless communication is increasingly being used for highly specific goal-oriented tasks. Designing systems which communicate as efficiently as possible under the constraints imposed by the specific task at hand will be important for optimizing the efficiency of wireless technologies, as well as improving performance of these technologies.

\section{A Different Approach: Context-Based Semantic Communication}
\label{sec_context}

Throughout this survey, we have presented a review of the history and state of the art of semantic communication by examining the different approaches toward engineering this higher level of communication. Each of these approaches differ in how they treat the semantics of the problem. Classical approaches attempt to quantify semantic information in probabilistic terms, much the same way as traditional information theory. KG-based semantic communication uses KG's to represent knowledge of the semantic source and receiver, from which semantic methods can be derived. In ML-based semantic communication, data is used to learn the latent ``semantic'' relationships and optimize communication based on these relationships. Significance-based semantic communication essentially combines ``The semantic problem'' with ``The effectiveness problem'' and emphasizes efficient, goal-oriented communication.

In this section, we present a novel approach to the semantic communication problem. We first argue that \textit{context} is at the heart of all semantic communications. While context is certainly an implicit consideration in each of the aforementioned approaches, we believe than an explicit and deliberate focus on the context of communication will lead to novel and valuable semantic communication systems. We then present our view of how to define context, and our vision of a systematic design procedure which frames semantic communication as a context-dependent, goal-oriented optimization problem.

\begin{figure*}[htbp]
    \centering
    \includegraphics[scale = 0.6]{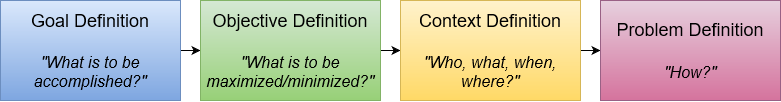}
    \caption{Proposed design flow for context-based semantic communication systems}
    \label{fig_design_flow}
\end{figure*}

\subsection{The Importance of Context}

To motivate the utility of a context-based approach, recall again the example from Section \ref{sec_intro}, where a speaker wants to communicate how to compute the area of a circle to a listener. Semantic communication in scenario 1 (listener vaguely familiar with geometrical concepts) can be much more \textit{syntactically} efficient than the same semantic communication in scenario 2 (listener is a small child). As was illustrated in Section \ref{sec_intro}, the key observation is that the listener in either scenario starts with a different prior knowledge base. 

However, it is important to note that this is not the \textit{only} characteristic of the scenario which will affect communication. What if scenario 1 takes place in a one-on-one office meeting, while scenario 2 takes place in a crowded classroom of restless children of similar age? Certainly the efficiency-of-communication gap between the two scenarios will widen. Furthermore, say that the speaker has a one-hour one-on-one office meeting with the listener in scenario 1, but has four 15-minute sessions in the crowded classroom with the young listener of scenario 2 held on different days. This again will impact how to most efficiently communicate in each scenario; it is likely that some review will be needed in each of the disjoint sessions of scenario 2.

Take any situation in which communication occurs, and a similar analysis can be done to determine the factors that impact the way in which communication is carried out. Based on this observation, the characterization of these factors is clearly an important step in designing an efficient semantic communication system. While context is inherent in any approach to semantic communication, the previously discussed methods only implicitly consider this key factor. Based on the above example, we argue that an explicit focus on context is needed for optimal semantic communication.

\subsection{Defining Context}

We broadly refer to any factors that impact \textit{how} one efficiently communicates as being part of the \textit{context} of the problem. In the example, one factor was the parties involved in the communication process. Another was the setting, i.e., a quiet office vs. a noisy classroom. The third factor involved temporal aspects of the situation, i.e., a long, uninterrupted session vs. short sessions spanning multiple days. We observe that these factors each correspond to difference pieces of the overall context (people, place, and time). We therefore postulate that context can be completely described by considering what are sometimes referred to as the ``five W's'':
\begin{itemize}
    \item \textit{Who}: Agents involved in the communication process
    \begin{itemize}
        \item Includes source(s), receiver(s), and/or other agents that are involved indirectly
    \end{itemize}
    \item \textit{What}: The mode of communication
    \begin{itemize}
        \item Could be text, speech, etc.
    \end{itemize}
    \item \textit{Where}: Qualities of the environment in which communication occurs
    \begin{itemize}
        \item Specifies channel characteristics
    \end{itemize}
    \item \textit{When}: Temporal aspects of the problem/environment
    \begin{itemize}
        \item Considers static vs. dynamic agents, mode, channel, and goal
    \end{itemize}
    \item \textit{Why}: The purpose of communication
    \begin{itemize}
        \item Defines what is to be achieved
    \end{itemize}
\end{itemize}

These aspects of context can in turn be incorporated into the mathematical model of the communication problem at hand. For example, the \textit{What} aspect dictates the space of signals or symbols available to the source for communication, the \textit{Where} aspect defines which channel model is to be used, and the \textit{Why} aspect may be some function which specifies quality of service or other desired outcomes. By explicitly considering \textit{Why} as part of the context, this is similar to significance-based communication in that it also addresses ``The effectiveness problem.'' Another term for this is \textit{goal-oriented communication}. Once a mathematical description of each aspect of the context is available, they can be incorporated into an optimization framework to achieve efficient semantic communication. This contextual knowledge may be available \textit{a priori}, or it may need to be learned online using modern data-driven techniques. In either case, just like the speaker in the example, an optimal communication strategy can be devised by explicitly taking these different aspects into consideration.

\subsection{Context-Based Design}

Using context as it is defined above, we propose a general design procedure for semantic communication systems. Note that we use the term ``context-based'' to imply that some optimization is beign carried out based on explicit consideration of the context. The procedure, illustrated in Figure \ref{fig_design_flow}, involves the systematic construction of an optimization problem and consists of the following steps:
\begin{enumerate}
    \item \textit{Goal definition}: First, the \textit{why} aspect of the context is determined. As is true in any engineering setting, we must first know what problem we are solving; what is it we are trying to accomplish? This will be a constraint in the overall optimization problem.
    \item \textit{Objective definition}: This step will determine exactly what is to be optimized, e.g., energy efficiency, spectral efficiency, etc. As the name implies, this will determine the objective function to be optimized.
    \item \textit{Context definition}: Now, define the remaining aspects of the context, namely \textit{who}, \textit{what}, \textit{where}, and \textit{when}. These will also manifest as constraints in the optimization problem.
    \item \textit{Problem definition}: Define a set of communication \textit{strategies} to be optimized over. Use this set, along with the objective function and constraints derived from steps 1-3, to define the optimization problem.
\end{enumerate}

As a general example, suppose we first obtain a goal represented by a constraint $C_G$. An objective function $f$ that we wish to minimize is identified, followed by $N$ context-based constraints $C_1, C_2, \ldots, C_N$. Finally, a set of possible communication strategies is determined and denoted by $\mathcal{S}$. Then, context-based semantic communication is performed by selecting a strategy as a solution to the optimization problem defined by
\begin{align*}
        \min_{S \in \mathcal{S}}.   \quad   & f(S)\\
        \textrm{s.t.}               \quad   & C_G, C_1, \ldots, C_N
\end{align*}

Depending on the characteristics of the resulting problem, how to solve it becomes a challenge in itself. If the resulting problem is convex, then well-known methods of solving are readily available \cite{boyd_convexopt}. In the likely case that the problem is non-convex, the problem becomes harder to solve, in which methods such as convex relaxation or ML may be needed to make the problem tractable. 

As a concrete example, consider smart agriculture, which is expected to play a prominent role in the 6G network \cite{alwis_survey6Gfrontiers_2021}. Furthermore, suppose we desire to accurately monitor soil moisture in the field using context-based semantic communication between a set of $J$ sensors and a single fusion node (FN). Following the framework above, the goal is to produce an accurate picture of the soil moisture throughout the field; mathematically, this can be expressed by a metric called \textit{confident information coverage} \cite{wang_confidentInfo_2013}:
\begin{equation*}
    \Phi(x) = \sqrt{\frac{1}{T}\sum_{t=1}^T (z^t(x) - \hat{z}^t(x))^2}
\end{equation*}
where $z^t(x)$ and $\hat{z}^t(x)$ are the actual and estimated soil moisture values at point $x$ and time $t$, respectively, and $T$ is the time period over which estimation takes place. We say that the field is ``completely confident information covered'' if $\Phi(x) < \epsilon$ for all $x$ in the field $\mathcal{X}$. We take this to be our goal, and correspondingly the first constraint of the optimization problem.

Next, we must define the objective. Suppose that we are interested in maximizing the lifetime of the sensor network, and thus minimizing the power of the sensors, as is typical in an IoT application. Suppose that associated with sensor $j$ is a set of sensing powers $\mathcal{P}_{\text{sense}}^{(j)}$ and transmit powers $\mathcal{P}^{(j)}_{TX}$. Intuitively, we assume that greater sensing power will produce more accurate sensing, and greater transmit power will produce higher quality communication. Then the total power of the sensors is given by
\begin{equation*}
    P = \sum_{j = 1}^J (P_\text{sense}^{(j)} + P_{TX}^{(j)}),
\end{equation*}
where $P_\text{sense}^{(j)} \in \mathcal{P}_{\text{sense}}^{(j)}$ and $P_{TX}^{(j)} \in \mathcal{P}^{(j)}_{TX}$ for all $j \in \{1,2,\ldots,J\}$. We can now also define a set of communication \textit{strategies} as $\mathcal{S} = \bigtimes_{j \in J} (\mathcal{P}_S^{(j)} \times \mathcal{P}_T^{(j)})$, and the objective becomes
\begin{equation*}
    \min_{S \in \mathcal{S}}   \quad   P = \sum_{j = 1}^J (P_\text{sense}^{(j)} + P_{TX}^{(j)}).
\end{equation*}

To define the context, we must consider the ``four W's'' listed above. We will use a state-space representation to define the context. To address the \textit{Who} question, let $\Omega_s = \{\omega_1, \omega_2, \ldots, \omega_J\}$ and $\Omega_r = \{\omega_r\}$ represent the states of the sensors and the FN, respectively. For example, $\omega_i$ could indicate whether sensor $i$ is online or offline, and $\omega_r$ might indicate whether the FN is receiving or computing. \textit{What} refers to the symbols used to communicate; e.g., under some strategies a sensor may use more precise quantization than others, resulting in longer symbols. \textit{Where} will encompass the channel effects between each sensor and FN, and can be represented by $\Omega_c = \{\omega_1, \omega_2, \ldots, \omega_{J}\}$. Taking the cross product of the individual state-spaces gives the overall state-space $\Omega = \{ \Omega_s \times \Omega_r \times \Omega_c\}$. Finally, the \textit{When} question is addressed by temporal changes in the state space. Assuming that we observe the state at discrete time instances, this can be expressed by representing the state-space as a function of time $\Omega[n]$, $n = 1,2,\ldots,\infty$.

Putting all of this together as the final step in the process, we arrive at the context-based optimization problem
\begin{align*}
        \min_{S \in \mathcal{S}}.   \quad   & P = \sum_{j = 1}^J (P_S^{(j)} + P_T^{(j)})\\
        \textrm{s.t.}               \quad   & \Phi(x) < \epsilon, \:\: \forall \: x \in \mathcal{X}\\
                                            & \Omega = \Omega[n].
\end{align*}
By framing communication in this optimization framework, the system is acting as the teacher from our initial example, namely by communicating with the underlying goal of efficiency, which is achieved by considering the context in which communication is taking place. As mentioned above, this formulation only presents us with a problem for which solving is another matter. As our goal here is to introduce the framework itself, we leave further study of this second stage of the problem for future work.

Semantic communication is regarded as a promising solution for improving the efficiency of communication systems. More so than the previously discussed techniques, the proposed context-based method is formulated with this specific aim in mind. By considering this aim at the outset, and taking into consideration the context of the communication problem, we believe that the resulting semantic communication systems will have the potential to advance the state of the art once more.

\section{Conclusion}
\label{sec_conclude}

The aim of this survey is to provide a comprehensive and clear picture of the current state of the emerging field of semantic communications. The push toward semantic communication systems is motivated by the explosion of global data traffic demand in recent years, and the intuitive benefits that can be achieved through efficient communications. Defining semantics is a non-trivial problem, and some approaches have emerged in the literature.

Classical semantic information-based approaches attempt to extend the ideas of information theory to capture the semantics of information, and are based on ideas of logical probabilities and truthlikeness. Two prominent theories are TWSI and TSSI, and truthlikeness-based approaches extend these theories. The key idea of this approach is to follow the path of classical Information Theory by first quantifying semantic information, and developing results based on this quantification.

KG-based semantic communication focuses on the aspect of a knowledge base at a semantic source and receiver, using a KG to model such knowledge bases. This approach follows from the extensive work revolving around the semantic web. By representing knowledge in a graph structure, semantic similarity measures can be devised and forms of reasoning can be performed. This form of structure and working with knowledge is the key driver of this approach.

ML-based semantic communications uses modern learning techniques to carry out semantic communication in a data-driven manner. This includes DL methods, which learn semantics through neural network structures, and RL techniques, which learn semantics with an action-reward framework. Unlike the classic and KG-based approaches, this approach assumes no formal structure of the semantics to be learned, and puts the burden of learning these semantics on the model itself. Consequently, meaning is captured by the tuned model parameters that are found as a result of data-driven training.

Finally, significance-based semantic communications take a goal-oriented approach and look to communicate in a way that best achieves the goal. Significance is quantified by metrics pertaining to different qualities of information, such as freshness and value. Semantic sampling is also a critical point of this approach, which seeks to generate only information that is pertinent to the task at hand. By addressing the effectiveness problem of communication, this approach circumvents the semantic problem and inherently assumes that the semantics will be addressed within the effectiveness solution.

Regarding the problem of data traffic reduction, various works in KG-based and ML-based semantic communication have demonstrated quantitative results illustrating the potential to address this growing issue for diverse applications and use-cases, from simple text-based speech to network operations recommendation systems. For the classical and significance-based approaches, these results are fewer. One conclusion that can be drawn upon examination of these various results, is the need for standard evaluation procedures across the different approaches. If efficient communication is to be one of the main goals of semantic communication, standard metrics capturing this idea should be chosen as the field develops to facilitate the comparison between competing methods.

For each approach, we provide some challenges and opportunities that could inspire future work in the field. For the classical approach, a clear opportunity lies in the further development of a theory of semantic information. The KG-based approach includes many existing methods which can be extended; particularly, scalability and learning methods are two areas which require further improvements. A major drawback of the ML-based approach is a lack of interpretability, and model-based DL and neurosymbolic AI are two potential solutions to this issue. The significance-based approach is the most recent of the four, and thus provides many exciting opportunities for future work, particularly in the development and study of significance-oriented metrics. The most important challenge to the success of semantic communication is the ability to define and work with ``meaning.'' We believe that future semantic communication systems will leverage techniques across the different approaches to optimize these systems, and thus we view future work corresponding to each approach as important to the overall field.

Furthermore, we advocate for a fifth approach to engineering semantic communication, namely context-based semantic communication. This approach places an emphasis on the context of the communication problem, which in turn impacts the strategy that leads to efficient communication. Based on the observation that humans naturally optimize communication as a result of the context, our approach involves the formulation of the strategy selection as an optimization problem, which can be solved using traditional or modern techniques. We demonstrate the details of this approach with a smart agriculture example based on a soil moisture monitoring application.

Realizing this higher level of communication is an exciting problem that presents a plethora of challenges and opportunities for future work. In the age of ever-expanding AI and ML, it is only natural that we apply this intelligence to communication systems to reap the benefits therein. Our hope is that this survey will prove to be a useful guide to anyone interested in the engineering of semantic communication.

\printbibliography

\begin{IEEEbiography}[{\includegraphics[width=1in,height=1.25in,clip,keepaspectratio]{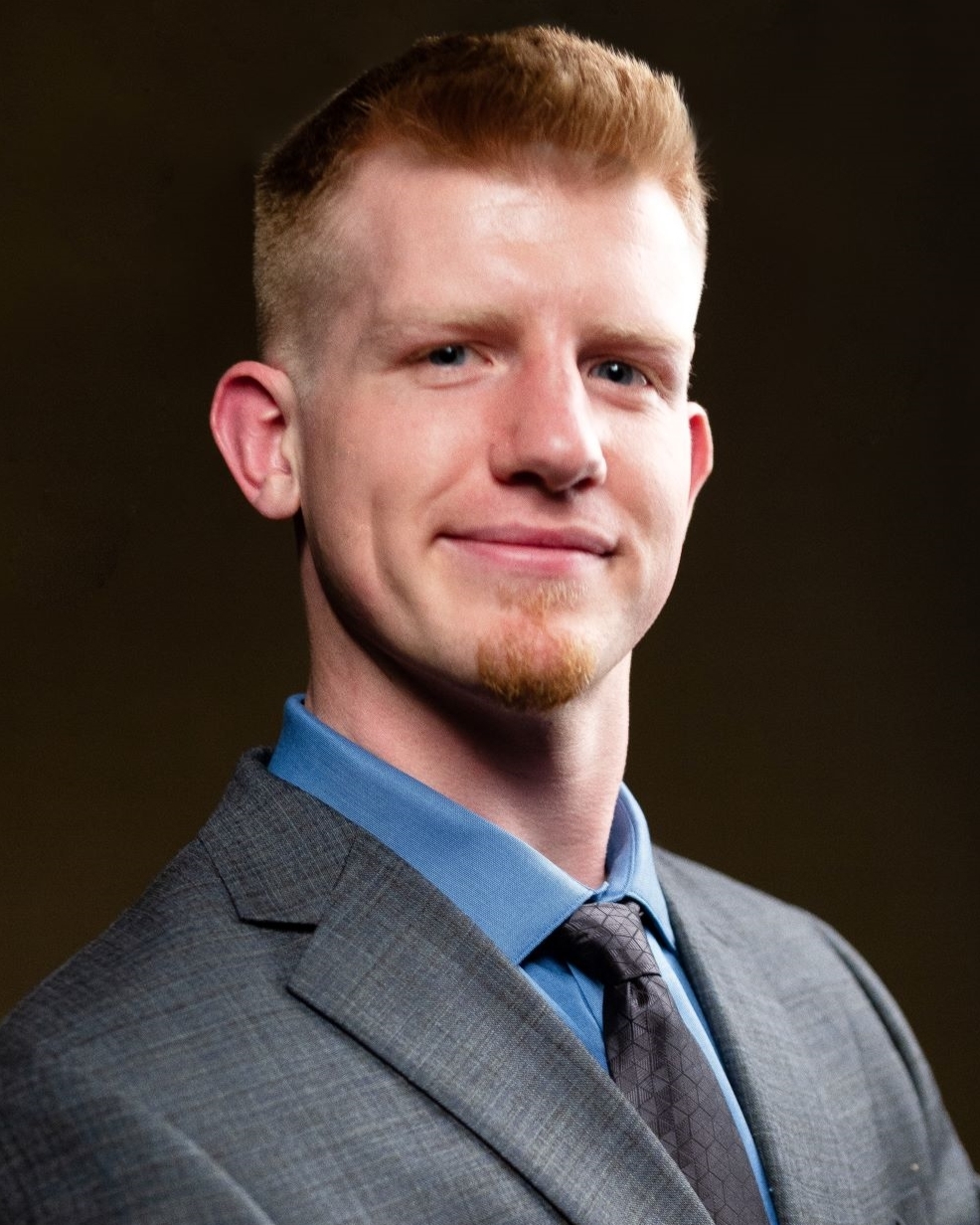}}]%
{Dylan Wheeler}
(Graduate Student Member, IEEE) received the A.S. degree from Highland Community College, Highland, KS, USA in 2016, the B.S. degree in Engineering from Ottawa University, Ottawa, KS, USA in 2018, and the M.S. degree in Electrical and Computer Engineering from Kansas State University, Manhattan, KS, USA in 2021. He is currently a Ph.D. student and a member of the Cyber-Physical Systems and Wireless Innovations Research Group at Kansas State University, Manhattan, KS, USA. His research interests include semantic communications, machine learning and artificial intelligence, and internet-of-things technologies for beyond-5G wireless networks.
\end{IEEEbiography}

\vskip 0pt plus -1fil

\begin{IEEEbiography}[{\includegraphics[width=1in,height=1.25in,clip,keepaspectratio]{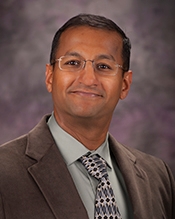}}]%
{Balasubramaniam Natarajan}
(Senior Member, IEEE) received the B.E. degree (Hons.) in electrical and electronics engineering from Birla Institute of Technology and Science, Pilani, India, Ph.D. degree in electrical engineering from Colorado State University, Fort Collins, CO, USA, Ph.D. degree in Statistics from Kansas State University, Manhattan, KS, USA, in 1997, 2002, and 2018, respectively. He is currently a Clair N. Palmer and Sara M. Palmer Endowed Professor and the Director of the Cyber-Physical Systems and Wireless Innovations Research Group. His research interests include statistical signal processing, stochastic modeling, optimization, and control theories. He has worked on and published extensively on modeling, analysis and networked estimation and control of smart distribution grids and cyber physical systems in general. He has published over 200 refereed journal and conference articles and has served on the editorial board of multiple IEEE journals including IEEE Transactions on Wireless Communications.
\end{IEEEbiography}

\end{document}